\documentclass[aps,pra,onecolumn,groupedaddress,tightenlines,nofootinbib]{revtex4}
\usepackage{amssymb}
\usepackage{color,graphicx}
\usepackage{lmodern}
\usepackage{amsmath}
\usepackage{amsbsy}
\usepackage{float}
\usepackage{bbm}
\usepackage{bm}
\usepackage{epsfig}
\usepackage{braket}
\usepackage{graphics}
\usepackage{empheq}
\usepackage{caption}
\usepackage{subfigure}
\usepackage{lipsum}
\usepackage{mathrsfs}
\usepackage{hyperref}
\usepackage{xcolor}
\usepackage{caption}

\hypersetup{
    colorlinks=true,
    linkcolor=blue,
    citecolor=blue,
    urlcolor=blue
}

\newcommand{\bq}{\begin{equation}} \newcommand{\eq}{\end{equation}}
\newcommand{\bqali}{\bq\begin{aligned}}
\newcommand{\eqali}{\end{aligned}\eq}
\newcommand{\bqn}{\begin{equation*}}
\newcommand{\eqn}{\end{equation*}}
\newcommand\D{\operatorname{d}}

\newcommand\I{\operatorname{I}}
\newcommand\HC{\operatorname{H.C}}

\newcommand\opt{\operatorname{opt}}

\begin{document}
\title{Spectral phase encoding for selective atomic excitation in quantum networks}
\author{Hamid Reza Naeij}
\email[]{hnaeij@mail.uni-paderborn.de}
\affiliation{Sharif Quantum Center, Sharif University of Technology, Tehran, Iran}

\begin{abstract}
Spectral phase encoding offers a means of distinguishing
photons addressed to different quantum-network nodes
within a shared optical bandwidth. We investigate how
this encoding controls the excitation of two-level atomic
receivers by single-photon wave packets. Using the
Heisenberg-Langevin formalism, we relate atomic excitation
to the temporal-mode overlap between the incident photon
and the time-reversed spontaneous-emission mode. Matched
spectral decoding restores the original photon mode and
its excitation response, whereas mismatched codes alter
the interference between spectral contributions and can
suppress unwanted excitation. We derive the mean cross-talk
for independently assigned random binary codes, accounting
for the frequency-dependent atomic response. When the
spectral contributions have comparable amplitudes and
phases, longer codes improve discrimination between
intended and unintended photons. We further determine
how random spectral-phase errors modify the time-dependent
atomic response and limit decoding fidelity. These results
connect spectral encoding and decoding to atomic-node
addressability and establish conditions for suppressing
cross-talk in multiuser quantum networks.
\end{abstract}

\maketitle

\section{Introduction}
Light-matter interactions provide a physical basis for
preparing, manipulating and detecting quantum states.
Their scope extends from the generation and preservation
of nonclassical optical states \cite{Lewenstein,Tirandaz}
to collective optical response \cite{Rui}, non-exponential
emission dynamics \cite{Andersson} and ultrastrong coupling
to electromagnetic continua \cite{Diaz}. Controlled
atom-light interactions also enable optical-state
discrimination \cite{Han} and single-photon routing
\cite{Shomroni}, while quantum optical fluctuations can
be exploited for random-number generation \cite{Moeini}.
These developments illustrate how the properties of light
and the response of matter can be combined to implement
quantum information tasks.

In quantum networks, this connection is particularly
important because propagating photons carry information
between spatially separated nodes, whereas localized
matter systems provide stationary degrees of freedom.
Photonic states can be transmitted through optical fibers
and free space and therefore serve as natural carriers
of quantum information \cite{Gheri,Flamini}. Their use
underlies approaches to quantum communication
\cite{Hu}, quantum computation \cite{Sheng2}, quantum
key distribution \cite{Sych1,Sych2} and entanglement
processing \cite{Sheng1}. Quantum teleportation and
hybrid information-processing schemes further emphasize
the importance of interfacing optical and material
degrees of freedom \cite{Furusawa}. For these interfaces,
a single photon must be described not only by its photon
number but also by its spectral and temporal mode,
which determines how it propagates, interferes, and
couples to a receiving system \cite{Brecht}.

The interaction of weak optical fields and single-photon
wave packets with individual emitters has consequently
received substantial attention. Relevant studies include
electromagnetically induced transparency with a single
atom \cite{Slodicka}, atomic excitation by exponentially
rising pulses \cite{Aljunid}, coupling to individual
molecules \cite{Wrigge}, plasmon-mediated light-matter
interactions \cite{Chang} and optical extinction by
single quantum dots \cite{Vamivakas}. For atomic
receivers, theoretical investigations have examined
spatial focusing \cite{Lindlein}, single-photon-induced
inversion in one-dimensional geometries \cite{Rephaeli},
coherent absorption in nanophotonic waveguides
\cite{Chen}, excitation by single- and multiphoton
wave packets \cite{Rag} and the influence of spatial
input modes \cite{Wang1}. Continuous-mode scattering
and excitation models provide a framework for connecting
these effects to the incident photon waveform
\cite{Domokos,Wang2,Tannoudji}. Together, these studies
establish that efficient excitation depends on matching
the incident optical mode to the spatial and temporal
response of the emitter.

The relevance of such mode matching is evident in recent
quantum-network demonstrations. Entanglement has been
distributed between nanophotonic quantum-memory nodes
connected by telecommunications fiber \cite{Knaut2024}
and trapped-ion networking protocols have been demonstrated
over an urban fiber link \cite{Kucera2024}. An interface
between photonic time-bin qubits and a solid-state
quantum-network node has also enabled quantum teleportation
using frequency conversion and waveform shaping to match
the spectral and temporal photon profiles
\cite{Iuliano2024}. Furthermore, near-unity photon
indistinguishability has been achieved between dissimilar
and independent atomic nodes \cite{Hoffet2024}.
These results highlight a common requirement:
photons generated, transmitted, and received by different
physical systems must have compatible optical modes.

Beyond matching individual links, spectral and temporal
control can support communication involving multiple
users and many optical modes. Spatiotemporal shaping
has been investigated for single photons in multiuser
quantum networks \cite{Huang2023}, while multimode
storage has been demonstrated in a telecom-wavelength
quantum memory \cite{Wei2024}. Developments in
frequency-bin quantum photonics provide further tools
for processing information in the frequency domain
\cite{Myilswamy2025}, complementing broader progress
in scalable photonic technologies \cite{Wang2025}.
A related approach is spectral phase encoding, developed
in coherent optical code-division multiple-access (CDMA)
communication \cite{Salehi}. In this approach, different
phase patterns are applied across a shared spectral band,
allowing optical signals to be distinguished through
their encoding and decoding masks rather than through
separate carrier frequencies.
\begin{figure}
\centering
\includegraphics[scale=0.22]{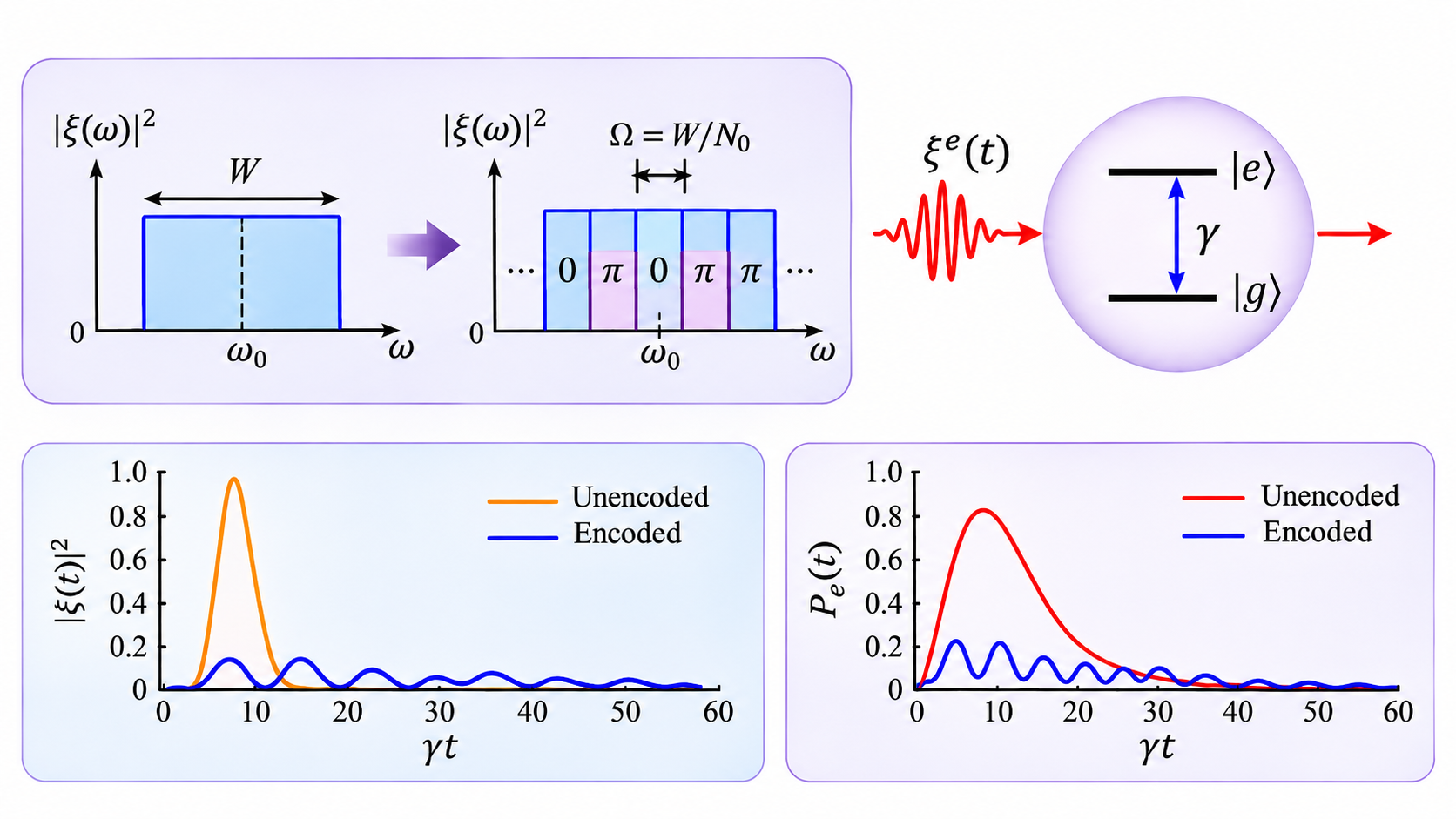}
\caption{Schematic illustration of the
interaction between a two-level atom and a
spectrally encoded single-photon wave packet.
Binary spectral phase encoding modifies the
relative phases of the photon frequency
components while preserving its spectral
intensity. The lower panels illustrate
the resulting temporal intensity redistribution
and the corresponding atomic excitation
dynamics, comparing encoded and unencoded
single-photon wave packets. The curves are
schematic representations of the physical
mechanism.}
\label{fig:schematic}
\end{figure}
Applying this principle to single-photon interfaces
raises a specific question: how does a spectral phase
code affect the excitation of an atomic receiver and
how accurately must that code be decoded to distinguish
an intended photon from other photons in the same band?
Spectral phase encoding preserves the photon number
and spectral intensity, but changes the interference
between frequency components and therefore the temporal
wave packet. Consequently, code discrimination at an
atomic receiver depends on both the residual spectral
phases and the frequency-dependent atomic response.
The connection between these quantities is needed to
assess node addressability, decoding tolerance, and
cross-talk within a common physical description.

In this work, we investigate this connection for a
two-level atom initially in its ground state and driven
by a spectrally encoded single-photon wave packet.
Using the Heisenberg-Langevin formalism, we calculate
the time-dependent excitation and relate it to the
overlap between the incident photon and the atomic
absorption mode. We examine how binary phase sequences
and spectral code length modify the photon waveform
and its excitation response. We then analyze matched
and imperfect decoding, derive the atomic response
averaged over random phase errors, and quantify
cross-talk between independently coded users through
spectral contributions weighted by the atomic response.
This formulation identifies the conditions under which
spectral encoding supports selective excitation of
quantum-network nodes and clarifies the limitations
imposed by imperfect mode matching and decoding.
Figure \ref{fig:schematic} illustrates the underlying
connection between spectral phase, temporal interference,
and atomic excitation.

The paper is organized as follows.
Section \ref{sec:model} introduces the atom-photon
interaction model and the temporal-mode overlap bound.
Section \ref{sec:encoded} examines spectral phase encoding
and its effect on atomic excitation.
Section \ref{sec:application} analyzes decoding errors,
multiuser cross-talk, and their implications for
quantum-network interfaces.
Section \ref{sec:conclusion} summarizes the results.

\section{Model}
\label{sec:model}
In this section, we employ the theoretical
framework developed in \cite{Domokos,Wang2} to describe the
dynamics of a two-level atom interacting
with a continuous-mode quantum optical field.
The atom-photon interaction is formulated
using the Heisenberg-Langevin equations,
which account for the atomic evolution
and its coupling to the surrounding
electromagnetic environment.

\subsection{Heisenberg-Langevin formalism for atom-photon interaction}
We consider the interaction of a two-level
atom with a propagating single-photon wave
packet, as schematically illustrated in
FIG. \ref{fig:schematic}. The atom is
characterized by a ground state $|g\rangle$
and an excited state $|e\rangle$, coupled
through an electric-dipole transition.
The incident photon is described by a
continuous superposition of frequency modes,
whose spectral distribution determines its
temporal wave-packet profile. To investigate
the resulting atom-photon dynamics, we
describe the electromagnetic field using
a continuum of quantized modes coupled
to the atomic transition. In the
interaction picture, the corresponding
quantized electric-field operator can
be expressed as \cite{Loudon}
\bq
E (t) =i \int \D \omega (\frac{\hbar\omega}{4\pi \epsilon_0 c A})^{1/2}\big[a(\omega)e^{-i\omega t}-a^\dagger(\omega)e^{i\omega t}\big] 
\eq
where $a^\dagger(\omega)$ and $a(\omega)$ are the creation and annihilation operators of the quantized field with the frequency $\omega$, respectively. The commutation relation of these operators is $\big[a(\omega),a^\dagger(\omega)\big]=\delta (\omega-\omega')$, where $\delta$ denotes the Dirac delta function. Moreover, $\epsilon_0$, $c$ and $A$ are the vacuum permittivity, the speed of light and the cross section, respectively.  

In the interaction picture and using rotating wave approximation (RWA), one can show that the interaction Hamiltonian of the system can be written as
\bq
H_{\I} =-i \hbar \int \D \omega \lambda_\omega \big[\sigma_+a(\omega)e^{-i(\omega-\omega_A) t}-\HC \big]
\eq
where $\omega_A$ is the atomic transition frequency and $\sigma_z=\vert e \rangle \langle e \vert-\vert g \rangle \langle g \vert$, $\sigma_+=\vert e \rangle \langle g \vert$, $\sigma_-=\vert g \rangle \langle e \vert $ are the atomic operators and H.C denotes Hermitian conjugates. Furthermore, the coupling strength is $\lambda_\omega=d \big(\omega/4\pi \epsilon_0 \hbar cA\big)^{1/2} $ where $d$ is the value of the dipole moment of the atom. 

The dynamics of the atomic operators are
governed by the following set of
Heisenberg-Langevin equations
\cite{Domokos,Wang2,Tannoudji}
\bqali
\label{eq:sigmaminus}
\frac{\D \sigma_-}{\D t}=-\gamma \sigma_-/2+\sigma_z \int \D \omega \lambda_\omega a(\omega)e^{-i(\omega-\omega_A) t}+\zeta_-
\eqali
\bq
\label{eq:sigmaz}
\frac{\D \sigma_z}{\D t}=-\gamma (\sigma_z+1)-2\int \D \omega \lambda_\omega \big[\sigma_+a(\omega)e^{-i(\omega-\omega_A) t}+\HC \big]+\zeta_z
\eq
where $\gamma$ is the decay rate of the atomic dipole and can be defined as $\gamma=2\pi\int \D \omega \vert \lambda_\omega\vert^2 \delta (\omega-\omega_A)$. Since we also considered the interaction between the atom and the environment, the terms of noise operators ($\zeta_z$ and $\zeta_-$) have appeared in the Heisenberg-Langevin equations, which are defined as 
\bqali
\zeta_-=\sigma_z\int \D \omega'\lambda_{\omega'} b (\omega')e^{-i(\omega'-\omega_A) t}
\eqali
\bqali
\zeta_z=&-2\int \D \omega' \lambda_{\omega' }\big[\sigma_+b(\omega')e^{-i(\omega'-\omega_A) t}+\HC \big]
\eqali
where $b^{\dagger} (\omega')$ and $b(\omega')$ are the creation and the annihilation operators of the environmental modes with the frequency $\omega'$, respectively. 
In the following analysis, we adopt the
narrowband approximation, assuming that
the spectral bandwidth $W$ of the
single-photon wave packet is much smaller
than its central frequency $\omega_0$.
Under this condition, the frequency dependence
of the atom-field coupling strength
$\lambda_\omega\propto\sqrt{\omega}$
can be neglected over the spectral range
occupied by the photon. Consequently,
the coupling strength can be approximated
by its value at the central frequency,
$\lambda_\omega\simeq\lambda_{\omega_0}
=d\big(\omega_0/4\pi\epsilon_0\hbar cA
\big)^{1/2}$.
This approximation allows us to treat
the coupling strength as frequency-independent
across the photon bandwidth while retaining
the spectral and temporal structure of
the incident wave packet.

\subsection{Excitation of a two-level atom by a rectangular-spectrum single-photon wave packet}
We now calculate the excitation probability $P_e(t)$ of a two-level atom in interaction with a single-photon wave packet which can be defined as \cite{Scully}
\bq
\label{eq:probability}
P_e(t)=\frac{1}{2} \big[ \langle \Psi_0 \vert\sigma_z (t) \vert \Psi_0 \rangle +1 \big]
\eq
We suppose the initial state of the total system is $\vert \Psi_0 \rangle=\vert g \rangle \vert 1_{\xi}\rangle \vert 0_e \rangle$ where is a product state of the atomic ground state, the single-photon wave packet state, and the vacuum state of the environment, respectively. 

A single-photon wavepacket state can be defined as $\vert 1_{\xi}\rangle=a^{\dagger}_{\xi} \vert 0 \rangle$ where 
\bq
\label{eq:single-photon}
a^{\dagger}_{\xi}=\int \D \omega \xi(\omega) a^{\dagger}(\omega) 
\eq
where $\xi(\omega)$ and $a^{\dagger}_{\xi}$ represent the spectral amplitude of the photon wave packet and the photon wave packet creation operator, respectively \cite{Loudon}. Moreover, we assume the environment to be initially in the vacuum state. Based on the Born approximation, the changes in the state of the environment are negligible \cite{Schlosshauer}. So, the state of the environment remains vacuum at all times. Considering the initial vacuum state of the environment, the average values of the noise operators in Eqs. \eqref{eq:sigmaminus} and \eqref{eq:sigmaz} will be zero. 
To obtain the excitation probability $P_e(t)$ of a two-level atom, we need to calculate the expectation value of the atomic operator $\sigma_z$ as $\langle \sigma_z (t) \rangle$ in Eq. \eqref{eq:probability}. So, one should solve the set of coupled differential equations for the atomic operators ($\sigma_z$ and $\sigma_{\pm}$) in Eqs. \eqref{eq:sigmaminus} and \eqref{eq:sigmaz}.  Moreover, we should define the explicit shape of the single-photon wave packet.

To investigate the atom-photon interaction,
we consider a single-photon wave packet
with a rectangular spectral profile
as an illustrative example \cite{Salehi}
\bqali
\label{eq:rectangular}
\xi(\omega)= \begin{cases} \sqrt{P_0}/W  & \mbox{for} -\frac{W}{2}\leq \omega \leq \frac{W}{2} \\
0  & \mbox{elsewhere}
\end{cases}
\eqali
where $P_0$ and $W$ are the peak power and the total bandwidth of the pulse, respectively. Therefore, one can show that the temporal shape pulse is a sinc function:
\bq
\label{eq:temporal}
\xi(t)=\sqrt{P_0}  \mbox{sinc} (\frac{W}{2} t) 
\eq
where $\mathrm{sinc}(x)=\sin(x)/x$.
For the numerical analysis, we consider
a photon spectral bandwidth $W=1.5\gamma$,
where $\gamma$ denotes the atomic decay rate.
This choice allows us to investigate the
excitation dynamics when the photon bandwidth
is comparable to the atomic linewidth.
Using the temporal profile in Eq. \eqref{eq:temporal},
we calculate the time-dependent excitation
probability $P_e(t)$ of a two-level atom
interacting with the rectangular-spectrum
single-photon wave packet. The resulting
dynamics serve as a reference for evaluating
the influence of spectral phase encoding
in Section \ref{sec:encoded}.
 
\subsection{Optimal temporal-mode matching and
atomic excitation bound}

The excitation probability of a two-level atom
depends not only on the spectral bandwidth of
the incident single-photon wave packet but also
on its temporal profile and its overlap with
the atomic absorption mode. For a two-level
atom coupled to a propagating optical mode
under the Markov approximation, maximum
single-photon absorption at a specified time
is achieved when the incident photon has
the time-reversed temporal profile of
spontaneous emission \cite{Stobinska}.
Physically, this corresponds to an
exponentially rising wave packet whose
temporal shape compensates for the natural
exponential decay of the atomic excitation,
allowing the photon to couple coherently
to the atomic transition. Taking $t=0$ as
the target excitation time, the corresponding
normalized optimal temporal mode is given by

\begin{equation}
\label{eq:temporal_profile}
\xi_{\opt}(t)= \sqrt{\gamma} e^{\gamma t/2} \Theta(-t)   
\end{equation}
where $\Theta(-t)$ is a Heaviside step function; the excitation probability achieved by an arbitrary normalized photon temporal mode $\xi(t)$ obeys the overlap bound
\begin{equation}
\label{eq:bound}
P_e^{\max}
=\max_t P_e(t)
=\beta\max_t
\left|
\int_{-\infty}^{\infty}
d\tau\,\xi^*(\tau)
\xi_{\mathrm{opt}}(\tau-t)
\right|^2
\leq\beta
\end{equation}
Here, $\beta\in[0,1]$ denotes the coupling
efficiency between the incident photon mode
and the atomic transition, while
$\xi_{\mathrm{opt}}(\tau-t)$ represents
the optimal absorption mode shifted to the
observation time $t$. The incident photon
is normalized according to
$\int d\tau\,|\xi(\tau)|^2=1$.
The maximum excitation probability is
determined by maximizing the temporal-mode
overlap between the incident photon and the
time-shifted optimal absorption mode.
The bound in Eq. \eqref{eq:bound} follows
from the Heisenberg-Langevin equations
\eqref{eq:sigmaminus} and \eqref{eq:sigmaz},
which, in the single-excitation manifold,
yield a convolution of the incident
photon amplitude with the atomic impulse
response
$h(t)=\sqrt{\beta\gamma}\,
e^{-\gamma t/2}\Theta(t)$,
where $\gamma$ is the total atomic decay
rate. Applying the Cauchy-Schwarz
inequality gives $P_e^{\max}\leq\beta$,
with equality attained when the incident
photon matches the time-shifted optimal
absorption mode. In our numerical simulations,
we assume ideal coupling ($\beta=1$),
so the maximum achievable excitation
probability is unity. The detailed
derivation of Eq. \eqref{eq:bound},
using the equivalent single-excitation
representation of the same atom-photon
dynamics, is provided in
Appendix \ref{app:derivation}.

\section{Spectrally encoded single-photon wave packets}
\label{sec:encoded}
In this section, we investigate the interaction
between a two-level atom and a spectrally encoded
single-photon wave packet, focusing on how spectral
phase modulation influences the atomic excitation
dynamics. We first introduce the spectral phase
encoding approach and then analyze the effects
of the encoding sequence and code length on the
temporal profile of the photon and the resulting
atomic excitation probability.

\subsection{Spectral phase encoding}
The spectral phase encoding approach to encode
the spectrum of a photon wave packet is used
in CDMA communication systems. See \cite{Salehi}
for a general review.
In this approach, the spectrum of a single-photon
wave packet is divided into several distinct
frequency intervals, referred to as spectral
chips, and a prescribed phase shift is applied
to each interval. In practice, these phase
shifts can be implemented using a programmable
spectral pulse shaper, which independently
controls the phase of different frequency
components. The resulting encoded photon
remains a single-photon state, with its
spectral intensity and total photon number
unchanged. However, the relative phases
introduced between the spectral components
modify their interference in the time domain,
thereby changing the temporal shape and
intensity distribution of the photon wave
packet. In particular, binary spectral
encoding assigns phase shifts of either
$0$ or $\pi$ to individual spectral chips,
corresponding to phase factors of $+1$ or
$-1$, respectively. Different sequences of these phase shifts
produce different temporal wave-packet profiles,
which in turn affect the interaction of the
photon with a two-level atom. 

To describe
this encoding process mathematically, we
introduce a spectral phase-shifting operator
$U$, which applies a frequency-dependent
phase $\theta(\omega)$ to the photon wave packet
and can be expressed as 
 \bq
U=e^{-i \sum_{\omega} \theta(\omega) a^{\dagger}(\omega)a(\omega)}=\prod_{\omega} U(\omega) 
 \eq
where $\theta(\omega)$ is the phase shift of the frequency $\omega$ on the spectral wave packet of the pulse $\xi(\omega)$ and $U(\omega)=e^{-i \theta(\omega) a^{\dagger}(\omega)a(\omega)}$. In the Heisenberg picture, one can show that:
\bq
\label{eq:unitary}
U a^{\dagger}(\omega)U^{\dagger}=a^{\dagger}(\omega) e^{-i \theta(\omega)}
\eq
Applying Eq. \eqref{eq:unitary} to the photon wave packet creation operator in Eq. \eqref{eq:single-photon}, we have $Ua^{\dagger}_{\xi} U^{\dagger}=a^{\dagger}_{\xi^e}$ where
\bq
\xi^e (\omega)=\xi(\omega) e^{-i\theta(\omega)}
\eq
in which superscript $e$ denotes the encoded photon wave packet. Moreover, the encoded wave packet in the temporal domain can be obtained by the Fourier transform as
\bq
\xi^e (t)=\frac{1}{\sqrt{2\pi}} \int \D \omega \xi(\omega) e^{-i\big(\omega t +\theta(\omega)\big)}
\eq
Therefore, spectral phase encoding modifies
the phase of the complex spectral amplitude
$\xi(\omega)$ while preserving its magnitude
$|\xi(\omega)|$ and hence the spectral
intensity. However, since the temporal
profile is determined by the Fourier
transform of the spectral amplitude, the
introduced relative phases modify the
interference between frequency components,
thereby reshaping the photon wave packet
in the time domain, as explicitly shown
in Eq. \eqref{eq:wave_packet}.

\subsection{Atomic excitation by a spectrally encoded single-photon wave packet}
To investigate the effect of spectral phase encoding
on atomic excitation, we consider the single-photon
wave packet with the rectangular spectral profile
introduced in Eq. \eqref{eq:temporal}. Its spectrum
$\xi(\omega)$ is divided into $N_0$ nonoverlapping
frequency intervals, referred to as spectral chips,
each having an equal bandwidth $\Omega=W/N_0$.
By assigning a phase shift to each spectral chip,
the relative phases of the frequency components
are modified while their spectral intensities
remain unchanged. Following the spectral phase
encoding approach \cite{Salehi}, the temporal
profile of the encoded single-photon wave packet
can be expressed as

\bqali
\label{eq:wave_packet}
\xi^e (t)=\frac{\sqrt{P_0}}{N_0} \mbox{sinc} \big(\frac{\Omega}{2}t\big) \sum_{n=-N}^N \mbox{exp}\big[-i (n \Omega t+\phi_n)\big]
\eqali
where the phase $\phi_n$ denotes the $n$th element of a code consisting of $N_0=2N+1$. Moreover, the term $\mbox{sinc} \big(\frac{\Omega}{2}t\big)$ is a real function that represents the temporal width of the encoded wave packet, and the exponential term shows a periodic behavior with period $T=2\pi/\Omega$ whose temporal shape depends on the code elements $\phi_n$. Note that if we have $\phi_n=0$ for all $n$, Eq. \eqref{eq:wave_packet} reduces to Eq. \eqref{eq:temporal}.
\begin{figure}
\centering
\begin{subfigure}[]{
\centering
\includegraphics[scale=0.43]{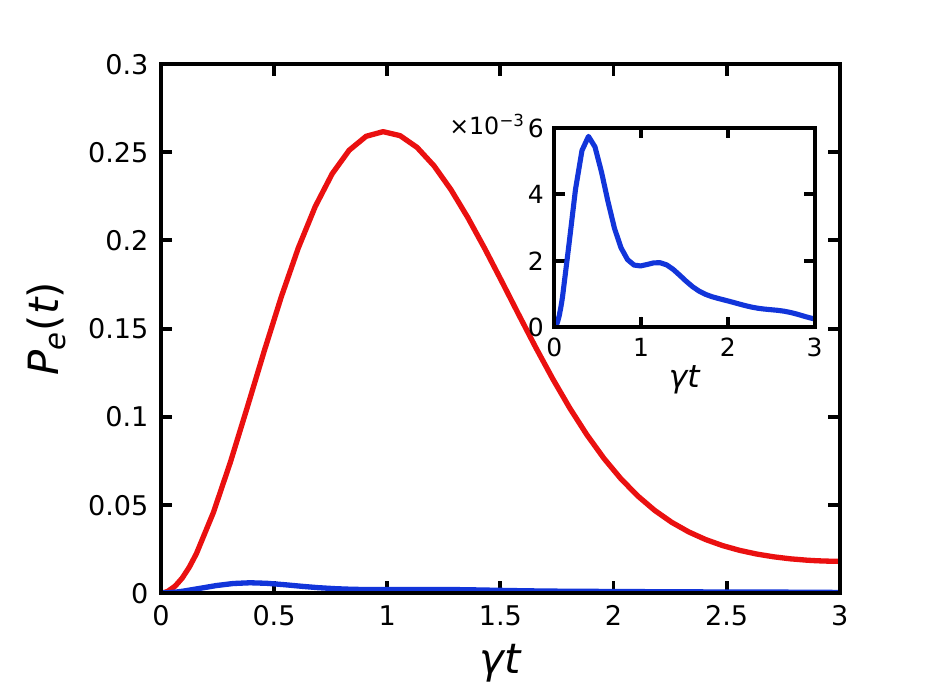}}
\end{subfigure}
\begin{subfigure}[]{
\centering
\includegraphics[scale=0.43]{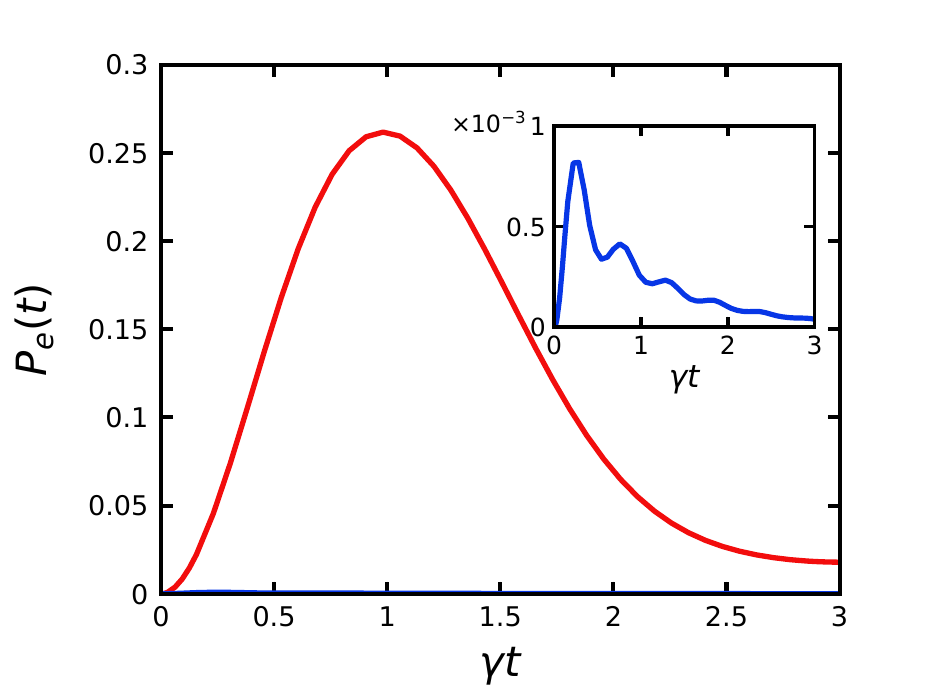}}
\end{subfigure} 
\begin{subfigure}[]{
\centering
\includegraphics[scale=0.43]{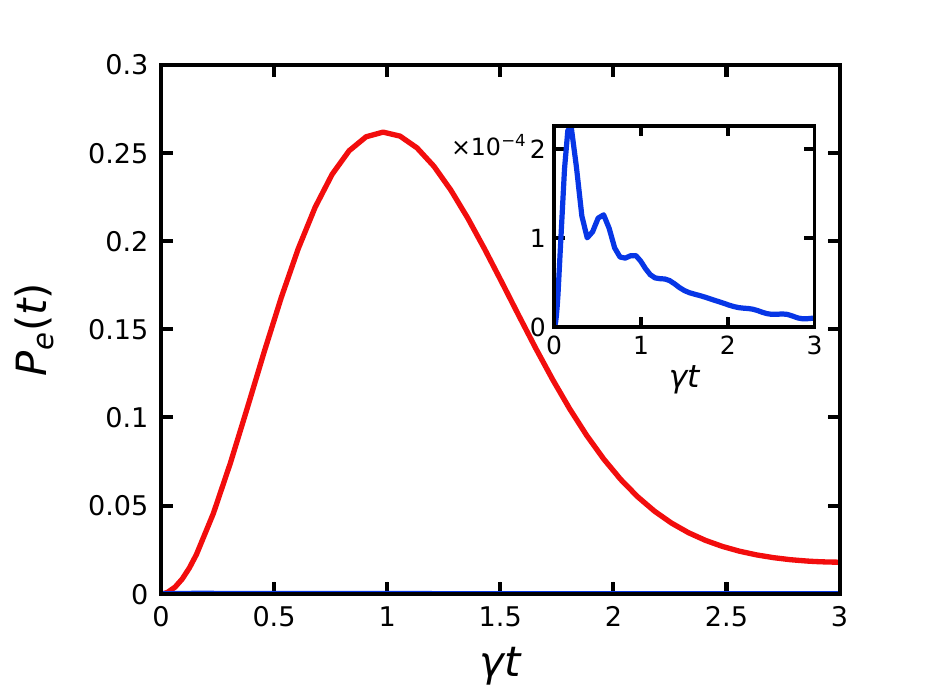}}
\end{subfigure}
\begin{subfigure}[]{
\centering
\includegraphics[scale=0.43]{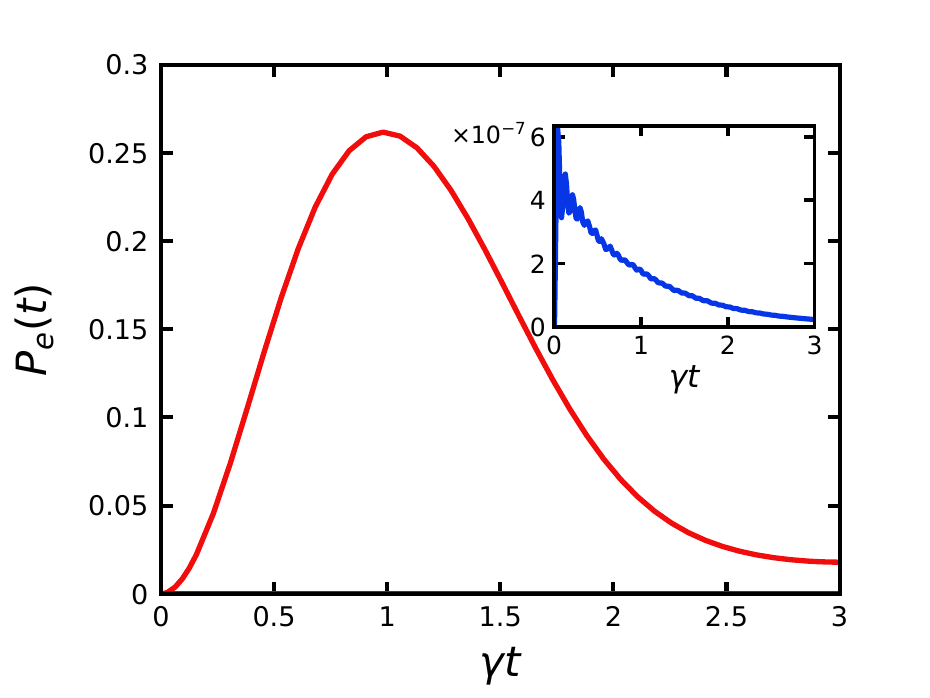}}
\end{subfigure}
\begin{subfigure}[]{
\centering
\includegraphics[scale=0.43]{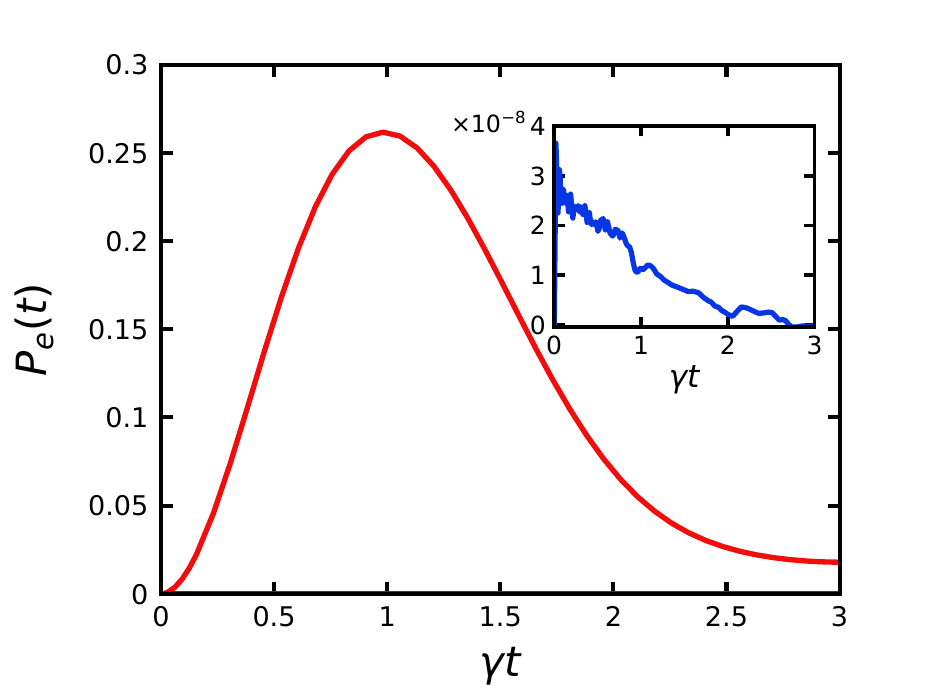}}
\end{subfigure} 
\caption{Time-dependent excitation probability
$P_e(t)$ of a two-level atom interacting with
spectrally encoded single-photon wave packets
with rectangular spectral profiles (blue curves),
compared with the unencoded reference
(red curves). The excitation probability is
plotted as a function of the dimensionless
time $\gamma t$ for different spectral code
lengths: (a) $N_0=3$, (b) $N_0=5$,
(c) $N_0=7$, (d) $N_0=31$, and
(e) $N_0=63$. The inset plots highlight
the excitation dynamics of the encoded
photons on an enlarged vertical scale.}
\label{fig:excitation_probability}
\end{figure}

We now consider the code lengths
$N_0=3,5,7,31,63$ to calculate the excitation
probability $P_e(t)$ of a two-level atom
interacting with an encoded single-photon
wave packet according to Eq. \eqref{eq:probability}.
These values are selected to investigate
the effects of spectral encoding across
different code lengths. The shorter codes
($N_0=3,5,7$) represent wave packets with
relatively few spectral chips, whereas
$N_0=31$ and $63$ allow us to examine
the behavior of longer encoding sequences.
At a fixed total bandwidth $W$, increasing
$N_0$ decreases the individual chip bandwidth
$\Omega=W/N_0$ and consequently broadens
the temporal envelope
$\mathrm{sinc}(\Omega t/2)$ in Eq. \eqref{eq:wave_packet}.
This explains the substantial difference
in the temporal extent of the encoded
wave packets for $N_0=7$ and $31$.
The detailed temporal structure also depends
on the selected binary phase sequence.
Thus, the chosen code lengths provide
representative examples of short and long
spectral encoding sequences rather than
indicating a distinct physical transition
between them.
Furthermore, we assume the value of the phase $\phi_n$ is $0$ and $\pi$, which correspond to the amplitude of $+1$ and $-1$, respectively. This spectral encoding process is called random binary spectral encoding \cite{Salehi}. 

In the binary spectral encoding, we intentionally choose odd $N_0$ so that a central chip is centered at $\omega_0$ (the atomic transition), yielding a symmetric index set $n \in \{-N,...,0,...,N\}$. This choice avoids placing $\omega_0$ in between two chips (the even-$N_0$ case), which tends to introduce a systematic $\cos (\Omega t/2)$ modulation in the time-domain envelope and can lead to strong destructive interference at $t=0$ for many random $\{0,\pi\}$. We verified numerically that even $N_0$ exhibits a larger fraction of near-vanishing overlaps with $\xi_{\opt}(t)$, consistent with the analytical discussion below. In other words, for even $N_0$, the resonance lies between the two central chips $n\pm \frac{1}{2}$. Writing the sum over half-integer indices introduces an extra factor $\cos(\Omega t/2)$ in $c_{\phi}(t)$, which suppresses the integrand near $t\approx0$ for random $\{0,\pi\}$ phases and thus yields a larger set of near-zero overlaps in Eq. \eqref{eq:bound}.

\begin{figure}
\centering
\includegraphics[scale=0.65]{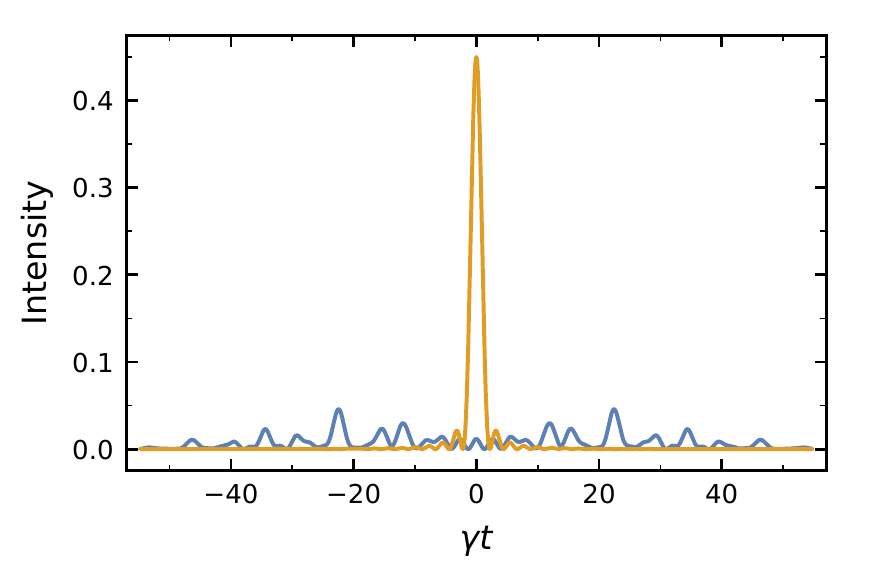}
\caption{Temporal intensity profiles of unencoded
(orange) and spectrally encoded (blue)
single-photon wave packets as functions of
$\gamma t$, for a binary spectral code of
length $N_0=31$. Spectral encoding redistributes
the photon intensity over a broader temporal
range, reducing its peak while preserving the
spectral intensity and photon normalization.}
\label{fig:intensity}
\end{figure}

We have plotted the excitation probability $P_e(t)$ of a two-level atom for the various code lengths of the encoded single-photon wave packet, which is compared to the unencoded wave packet in FIG. \ref{fig:excitation_probability}. We conclude that the encoding process of the single-photon wave packet decreases the excitation probability of a two-level atom. To analyze this result, we have plotted the intensity of the unencoded and encoded wave packet in FIG. \ref{fig:intensity}. As is clear in this figure, the encoding process causes the single-photon wave packet to spread in the time domain. Consequently, the power of the wave packet diminishes, and the wave packet behaves as a low-intensity pulse. Therefore, an encoded wave packet excites a two-level atom much weaker than an unencoded one.

\begin{figure}
\centering
\begin{subfigure}[]{
\centering
\includegraphics[scale=0.45]{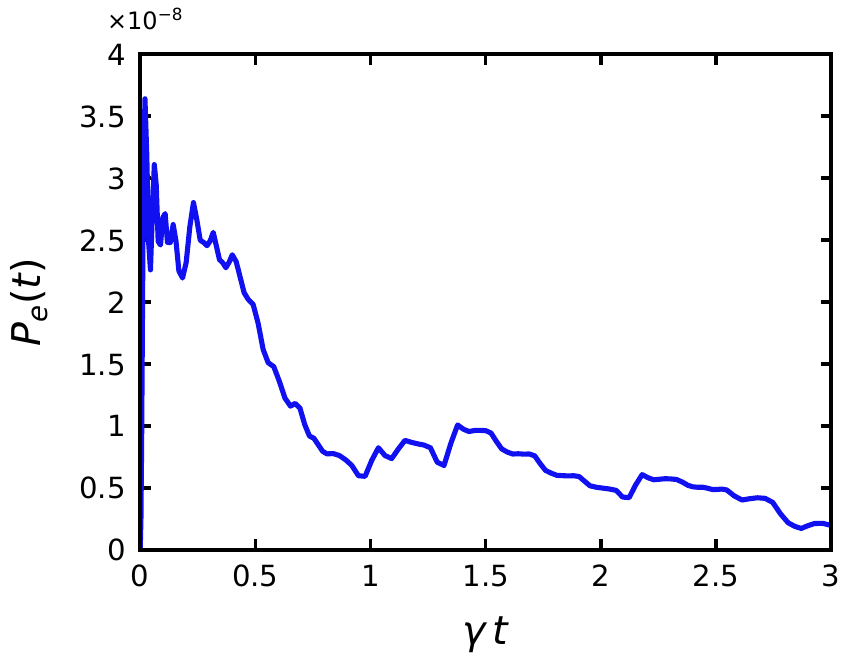}}
\end{subfigure}
\begin{subfigure}[]{
\centering
\includegraphics[scale=0.45]{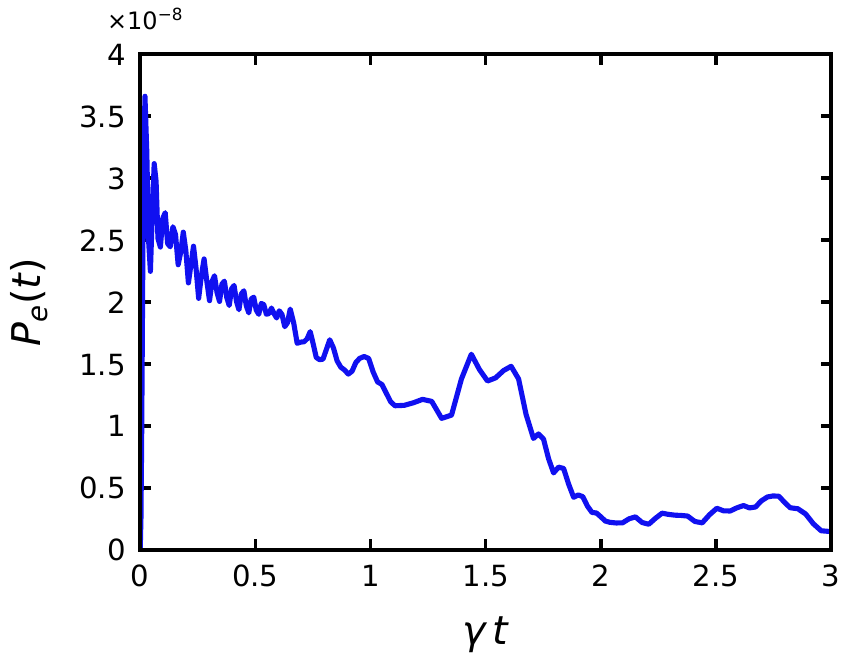}}
\end{subfigure} 
\begin{subfigure}[]{
\centering
\includegraphics[scale=0.45]{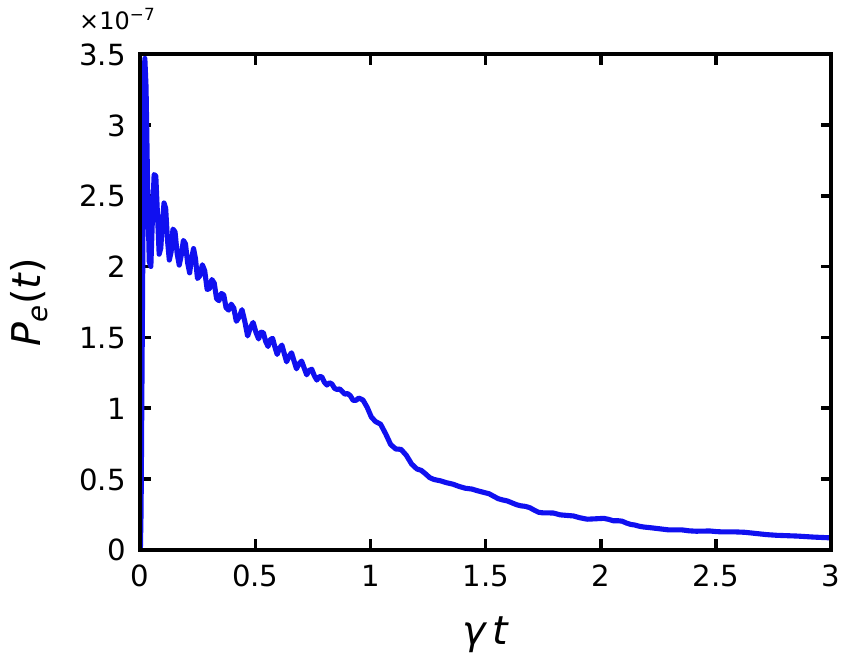}}
\end{subfigure}
\caption{Time-dependent excitation probability
$P_e(t)$ of a two-level atom interacting with
spectrally encoded single-photon wave packets
for three different random binary phase
sequences ($0$ and $\pi$), with a fixed
code length $N_0=63$. The distinct oscillatory
patterns illustrate the dependence of the
atomic excitation dynamics on the selected
spectral phase sequence.}
\label{fig:probability}
\end{figure}

To understand the origin of the oscillations
observed in the atomic excitation probability,
we examine how the binary spectral phase
sequences influence the temporal dynamics
of the atom-photon interaction. According
to Eq. \eqref{eq:wave_packet}, assigning
different phase shifts of $0$ or $\pi$ to
the spectral chips modifies the interference
between the frequency components of the
encoded photon, resulting in distinct
temporal wave-packet profiles.
FIG. \ref{fig:probability} illustrates
the excitation probability for three
different random binary phase sequences
with a fixed code length $N_0=63$.
The results demonstrate that the oscillatory
behavior of $P_e(t)$ depends on the specific
phase sequence, as different spectral
phase arrangements modify the temporal
interference pattern and, consequently,
the time-dependent coupling between the
photon and the atomic transition.

\begin{figure}[h]
\centering
\includegraphics[scale=0.6]{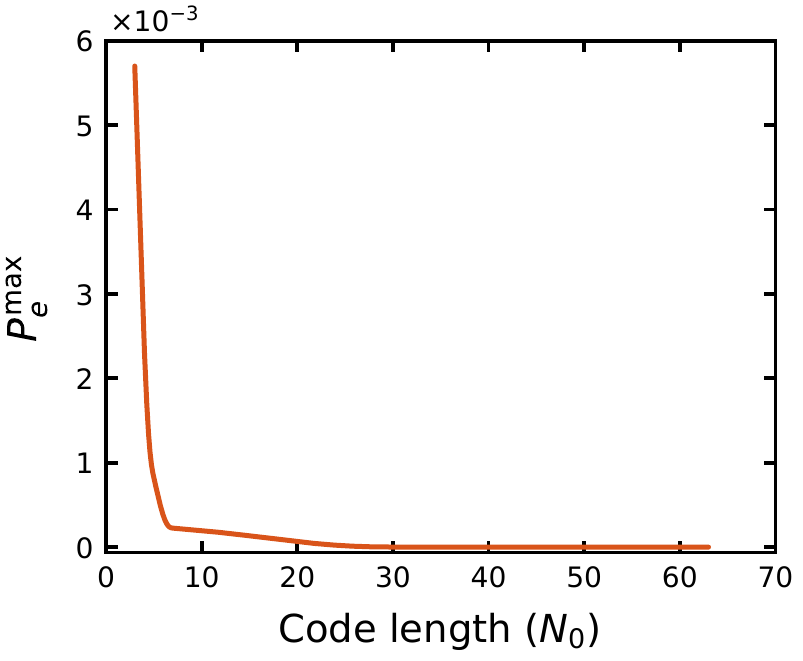}
\caption{Dependence of the maximum atomic
excitation probability $P_e^{\max}$ on the
spectral code length $N_0$ for binary
spectrally encoded single-photon wave packets.
The results show a decrease in the maximum
excitation probability with increasing code
length for the considered phase sequences,
reflecting the influence of spectral phase
encoding on temporal-mode matching between
the photon and the atomic transition.}
\label{fig:codelength}
\end{figure}

\begin{figure}[h]
\centering
\begin{subfigure}[]{
\centering
\includegraphics[scale=0.45]{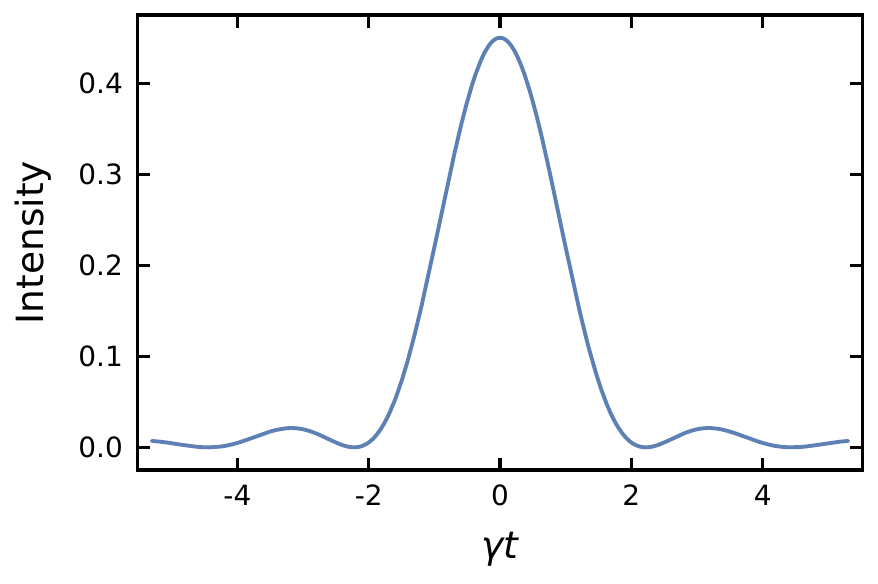}}
\end{subfigure}
\begin{subfigure}[]{
\centering
\includegraphics[scale=0.45]{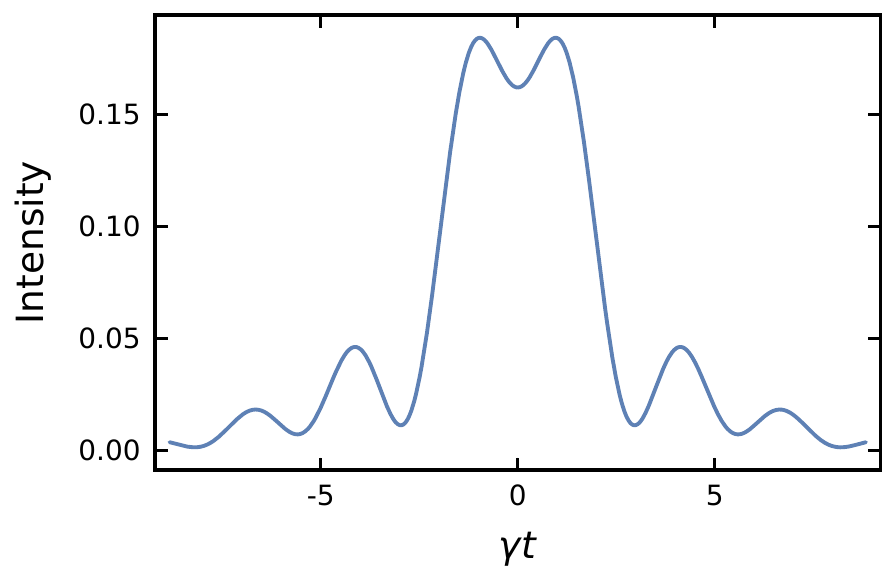}}
\end{subfigure} 
\begin{subfigure}[]{
\centering
\includegraphics[scale=0.45]{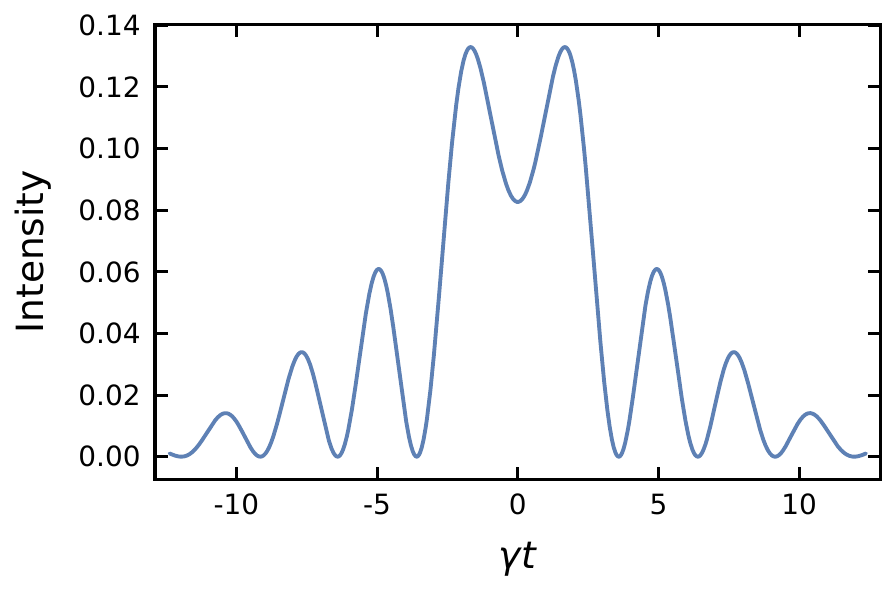}}
\end{subfigure}
\begin{subfigure}[]{
\centering
\includegraphics[scale=0.45]{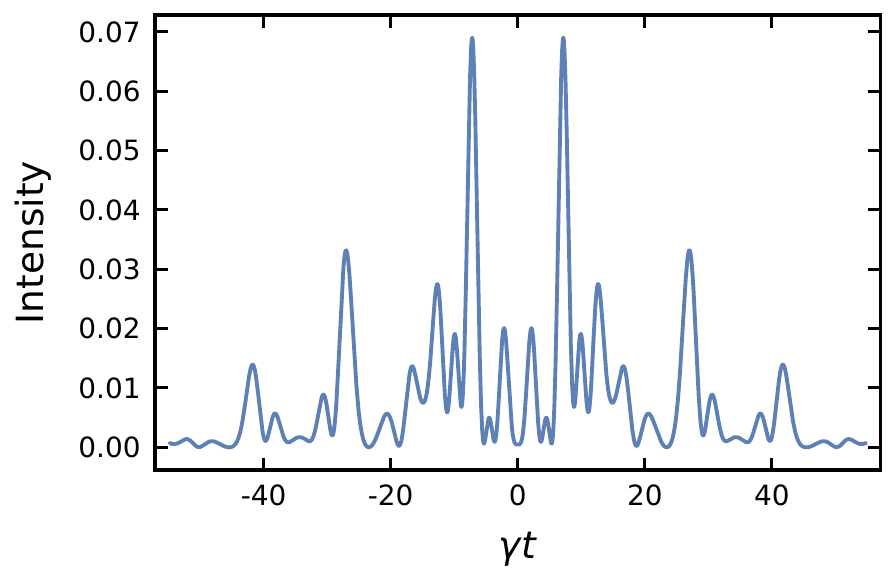}}
\end{subfigure}
\begin{subfigure}[]{
\centering
\includegraphics[scale=0.45]{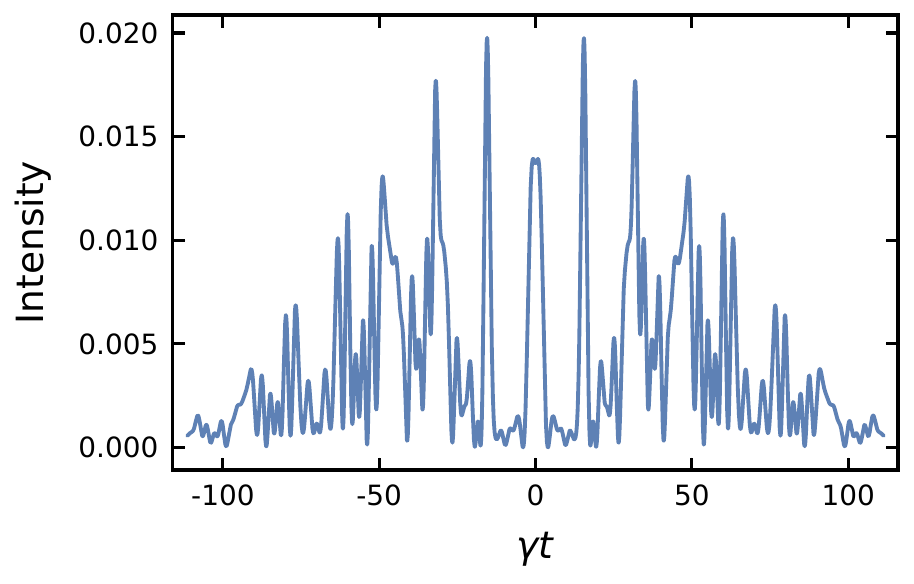}}
\end{subfigure}
\caption{Temporal intensity profiles of
spectrally encoded single-photon wave packets
for different binary code lengths:
(a) $N_0=3$, (b) $N_0=5$, (c) $N_0=7$,
(d) $N_0=31$, and (e) $N_0=63$.
The profiles are plotted as functions of
the dimensionless time $\gamma t$.
Increasing the code length broadens the
temporal envelope and redistributes the
photon intensity, while the detailed
temporal structure depends on the
selected spectral phase sequence.}
\label{fig:Spreading}
\end{figure}

We further investigate the influence of the spectral
code length $N_0$ on the maximum excitation
probability of a two-level atom, defined as
$P_e^{\max}(N_0)=\max_{t\geq 0}P_e(t;N_0)$.
FIG. \ref{fig:codelength} shows the dependence
of $P_e^{\max}$ on the code length for the
considered binary spectral phase sequences.
The results indicate that the maximum atomic
excitation probability decreases with increasing
code length. At a fixed total bandwidth $W$,
increasing $N_0$ reduces the individual spectral
chip bandwidth $\Omega=W/N_0$, thereby broadening
the temporal envelope of the encoded photon
wave packet, as illustrated in
FIG. \ref{fig:Spreading}. The resulting
redistribution of the photon's temporal intensity,
together with interference between its spectral
components, reduces the temporal-mode overlap
with the atomic response for the considered
phase sequences. Consequently, the maximum
excitation probability decreases, demonstrating
how the spectral code length influences the
efficiency of the atom-photon interaction.

To further understand how the spectral code
length and the selected binary phase sequence
influence the atomic excitation dynamics,
we examine the temporal structure of the
encoded single-photon wave packet.
As discussed above, increasing the code length
modifies the temporal distribution of the
photon, while different phase sequences
produce distinct interference patterns
between its spectral components.
These effects can be understood more clearly
by separating the temporal profile in
Eq. \eqref{eq:wave_packet} into two contributions:
a slowly varying envelope, which determines
the overall temporal extent of the wave packet,
and a phase-dependent sum, which describes
the interference between the encoded frequency
components. Accordingly, the encoded temporal
profile can be written as
$\xi^e(t)=s(t)c_\phi(t)$, where
$s(t)=\frac{\sqrt{P_0}}{N_0}
\mathrm{sinc}\left(\frac{\Omega t}{2}\right)$
is the temporal envelope, and the
phase-dependent contribution is defined as

\begin{equation}
c_{\phi}(t)=\sum_{n=-N}^N \exp [-in\Omega t+i\phi_n]   
\end{equation}
so that $\xi_e(t) \propto s(t) c_{\phi}(t)$. The overlap in Eq. \eqref{eq:bound} becomes
\begin{equation}
 \langle \xi_e \mid \xi_{\opt} \rangle \propto\int \D t \mbox{sinc} (\frac{\Omega t}{2}) e^{\gamma t/2} \Theta(-t)\sum_{n=-N}^N \exp [in\Omega t-i\phi_n]  
\end{equation}
Evaluating the time integral term-wise yields a closed-form dependence on $\Omega/\gamma$ and on the code phases $\{\phi_n\}$ via a weighted discrete Fourier transform of the sequence $\{e^{-i\phi_n}\}$. Consequently, (i) increasing the code length $N_0$ broadens $c_{\phi}(t)$ and reduces the overlap, explaining the monotonic trend of FIG. \ref{fig:codelength}; and (ii) different phase sequences produce distinct beating patterns in $P_e$ harmonics of $\Omega$ consistent with the oscillations seen in FIG. \ref{fig:probability}.

The temporal reference used in the figures
requires clarification. The photon wave
packets in Eqs. \eqref{eq:temporal} and
\eqref{eq:wave_packet} are defined over
both negative and positive times, with
$t=0$ corresponding to the center of the
unencoded pulse envelope. Accordingly,
FIGs. \ref{fig:intensity} and
\ref{fig:Spreading} display their complete
temporal intensity profiles. In contrast,
FIGs. \ref{fig:excitation_probability}
and \ref{fig:probability} describe a
finite-time initial-value problem in which
the atom is prepared in its ground state
and the atom-photon interaction is
initialized at $t=0$. The excitation
dynamics are therefore evaluated for
$t\geq 0$, with only the subsequent
portion of the incident wave packet
contributing to the atomic response.
For a continuously coupled atom, the
negative-time portion of the photon
would also contribute to excitation,
as described by the causal convolution
in Appendix \ref{app:derivation}. Thus, the different
temporal ranges reflect distinct
initial conditions rather than
different definitions of the photon
wave packet.

\section{Spectral encoding in quantum networks}
\label{sec:application}
\begin{figure}
\centering
\includegraphics[scale=0.27]{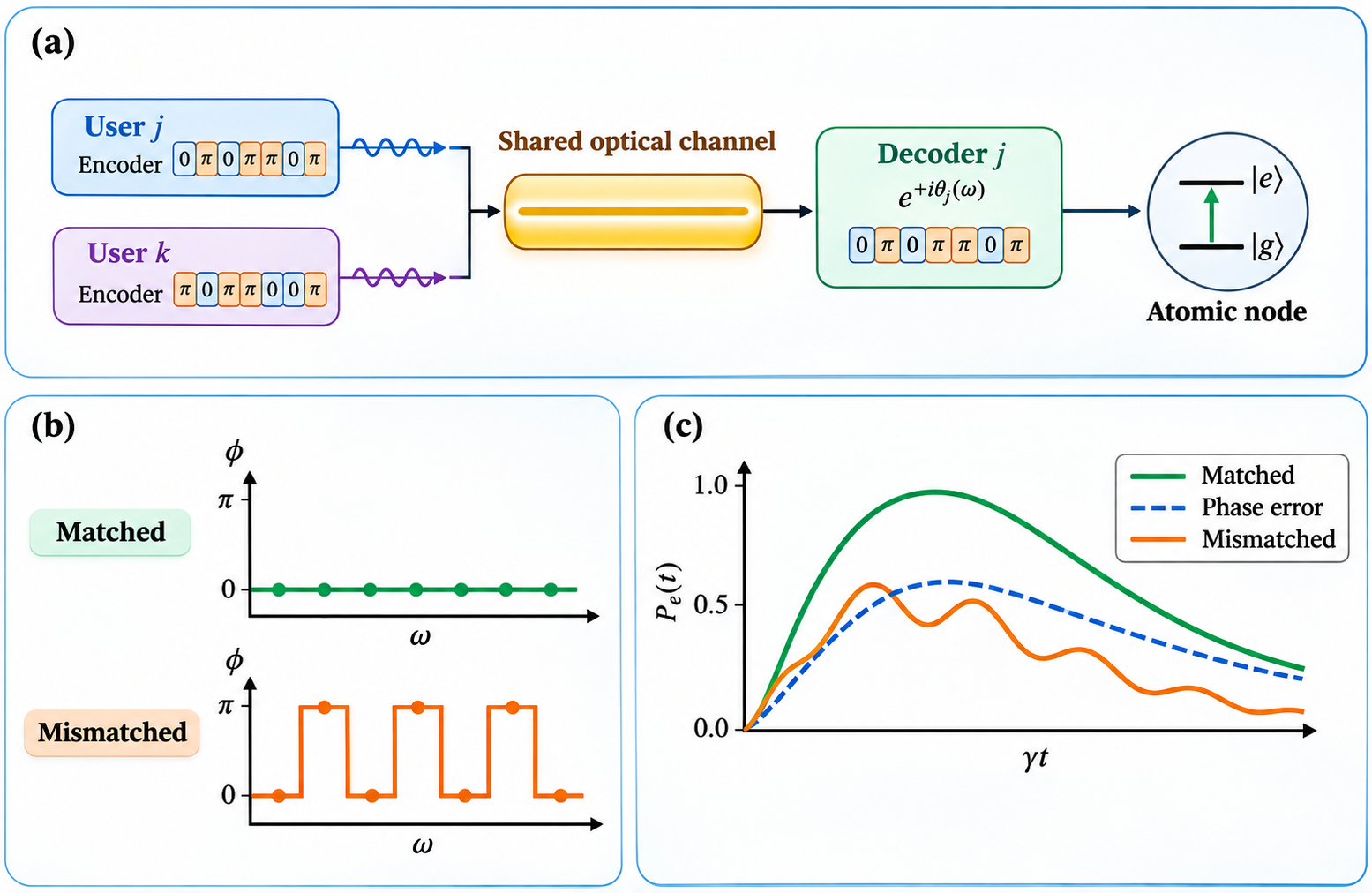}
\caption{Schematic illustration of spectral
encoding and mode-selective atom-photon
interactions in a quantum network.
(a) Multiple users transmit spectrally encoded
single photons through a shared optical channel
to an atomic receiver equipped with a
user-specific spectral decoder.
(b) Matched decoding cancels the imposed
spectral phases, whereas mismatched decoding
leaves residual phase differences between
the frequency components.
(c) Illustrative atomic excitation dynamics
for matched decoding, imperfect decoding,
and mismatched spectral codes.
The excitation curves are schematic and
do not represent numerical simulation results.}
\label{fig:quantum_network}
\end{figure}

A quantum network consists of spatially
separated quantum nodes connected through
quantum communication channels, enabling
the transfer of quantum information and
the distribution of entanglement between
distant systems. In such networks, photons
typically serve as carriers of quantum
information, while atoms or other quantum
emitters can act as stationary nodes for
storing and processing quantum states.
Efficient quantum communication therefore
requires controlled interactions between
propagating photons and individual quantum
nodes. In this context, the spectral
encoding approach investigated in this
work provides a possible mechanism for
controlling atom-photon coupling through
the spectral-temporal structure of
single-photon wave packets. In this section,
we examine the effects of spectral decoding,
phase errors, and multiuser cross-talk,
and discuss their implications for
selective node addressability and
quantum-network performance. 
FIG. \ref{fig:quantum_network} schematically
illustrates the role of spectral encoding and decoding in a multiuser quantum network,
highlighting how matched and mismatched
spectral codes influence the coupling of
single-photon wave packets to an atomic
receiver.

\subsection{Spectral phase decoding and mode matching}
In a spectrally encoded quantum communication
link, the interaction between an incoming
single photon and an atomic receiver depends
on the relative phases of its spectral
components. The encoder imposes a
frequency-dependent phase mask
$\theta(\omega)$, transforming the original
photon spectrum into
$\xi^e(\omega)=\xi(\omega)e^{-i\theta(\omega)}$.
Although this operation preserves the
spectral intensity and photon number,
it modifies the interference between
frequency components and consequently
reshapes the temporal wave packet arriving
at the receiving node. To recover the
original photon mode, the receiver applies
a compensating spectral phase mask
$D(\omega)=e^{i\tilde{\theta}(\omega)}$,
which can be implemented using a programmable
$4f$ pulse shaper. The resulting spectral
amplitude incident on the atom is
$\xi_{\mathrm{in}}(\omega)=
\xi(\omega)e^{i[\tilde{\theta}(\omega)
-\theta(\omega)]}$.
We define the residual phase mismatch as
$\phi(\omega)=\tilde{\theta}(\omega)
-\theta(\omega)$. For perfect decoding,
$\phi(\omega)=0$, and the original
spectral-temporal mode is recovered,
restoring its overlap with the atomic
absorption mode. In contrast, imperfect
decoding leaves frequency-dependent
phase differences that alter the temporal
interference pattern and generally reduce
the coherent coupling of the photon to
the atomic transition. The resulting
excitation probability therefore depends
not only on the photon spectral intensity
but also on how accurately its spectral
phase is compensated at the receiver.

Within the narrowband and Markov
approximations introduced in
Secion \ref{sec:model}, the frequency-dependent
response of the two-level atom is
characterized by a Lorentzian susceptibility
centered at the atomic transition frequency
\cite{Scully}. This response reflects the
finite linewidth associated with spontaneous
emission and determines how efficiently
different spectral components of the
incident photon couple to the atomic
transition. The corresponding atomic
susceptibility can be expressed as \cite{Scully}
\begin{equation}
\label{eq:Lorentzian}
\chi_A(\omega)=\frac{\sqrt{\gamma}}{\gamma/2-i(\omega-\omega_A)}
\end{equation}
where $\gamma$ is the dipole decay rate and $\omega_A$ is the atomic transition frequency. In the single-photon, weak-excitation regime used throughout, the excitation amplitude is proportional to the scalar product of the input mode with $\chi_A$
\begin{equation}
\label{eq:amplitude}
\mathcal{A} \propto \int \D\omega \xi(\omega)e^{i\phi(\omega)}\chi_A(\omega)
\end{equation}
We define the spectral-overlap functional \cite{Fiechtner}
\begin{equation}
\mathcal{M}[\phi,\Delta]=\frac{\int \D\omega \xi(\omega)e^{i\phi(\omega)}\chi_A(\omega-\Delta)}{\Big(\int \D\omega \mid \xi(\omega)\mid^2\Big)^{1/2}\Big(\int \D\omega \mid \chi_A(\omega)\mid ^2\Big)^{1/2}} 
\end{equation}
with dtuning $\Delta=\omega-\omega_A$. The spectral-overlap functional was defined by inserting the photon spectrum $\xi(\omega)$
 and the atomic susceptibility $\chi_A(\omega)$ into the excitation amplitude integral and normalizing it to be dimensionless. It is simply the normalized inner product between the photon mode and the atomic response. This functional directly measures how well the photon’s spectral-temporal mode matches the atomic absorption mode, and the excitation probability satisfies $P_e^{\max} \propto \mathcal{M}[\phi,\Delta]\mid^2$. Then, the peak excitation probability obeys the bound in Eq. \eqref{eq:bound}
\begin{equation}
P_e^{\max}  \leq\beta \kappa \mid \mathcal{M}[\phi,\Delta]\mid^2 
\end{equation}
where $\beta \in[0,1]$ is the coupling efficiency into the atomic mode (free-space or waveguide) and $\kappa$ is a geometry-dependent constant (absorbed in our normalization in Section \ref{sec:model}). Equality is approached when temporal mode matching is optimal. This recovers our observed suppression for encoded, undecoded photons, and quantifies recovery under correct decoding $(\phi=0)$.

\subparagraph{Bandwidth matching} For our rectangular spectrum of width $W$ in Eq. \eqref{eq:rectangular}, the perfectly decoded case $(\phi=0)$ yields
\begin{equation}
\mid \mathcal{M}[0,\Delta]\mid ^2 \approx \frac{1}{N} \mid \int_{-W/2}^{W/2}\frac{\gamma \D \omega}{\gamma^2/4+(\omega-\Delta)^2}  \mid ^2  
\end{equation}
with $N=\int \D \omega \mid \xi(\omega)\mid ^2 $. This increases with $W/\gamma$ up to the point where the Lorentzian tails limit further gains; for a given $\gamma$ there is an optimal $W$ balancing spectral weight near resonance against total pulse energy. This explains why increasing code length $N_0$ (fixed $W$) does not improve excitation unless decoding removes the imposed phase structure.

Our analysis is based on a two-level atomic
system within the Markov approximation,
characterized by a single total decay rate
$\gamma$, while neglecting atomic recoil
and additional internal energy levels.
The achievable excitation probability is
limited by the coupling efficiency
$\beta\leq 1$, as expressed in
Eq. \eqref{eq:bound}, and further depends
on the spectral-temporal overlap between
the incident photon and the atomic absorption
mode. More complex configurations, including
structured electromagnetic reservoirs,
non-Markovian environments, and multilevel
$\Lambda$-type systems, may modify the
atomic response and the optimal absorption
mode. Nevertheless, the underlying principle
of temporal-mode matching remains relevant:
efficient atom-photon coupling requires
the decoded photon wave packet to closely
match the absorption mode of the receiving
quantum node.

Furthermore, spectral phase encoding and
ideal decoding are unitary transformations
acting on the single-photon Hilbert space.
These operations preserve the photon number
and spectral intensity without measuring,
copying, or amplifying the quantum state.
Although unitary transformations preserve
inner products between equally transformed
states, spectral encoding can modify the
overlap of the photon with a fixed atomic
absorption mode. Applying the appropriate
inverse phase transformation restores the
original photon mode, whereas imperfect
or mismatched decoding generally reduces
the coupling efficiency through residual
spectral phase differences. These unitary
properties apply to the encoding and decoding
operations themselves, while the subsequent
atom-photon interaction may involve
absorption and dissipation. Consequently,
the proposed spectral encoding and decoding
scheme remains fully consistent with the
no-cloning theorem.

\subsection{Spectral phase mismatch and decoding tolerance}
\label{sec:tolerance}
Imperfect spectral decoding introduces
frequency-dependent phase errors that
modify the temporal interference between
the spectral components of the incident
single photon without changing its spectral
intensity. These residual phases reduce
the coherent overlap between the decoded
photon wave packet and the atomic absorption
mode, thereby suppressing the achievable
excitation probability. To quantify this
effect, we consider weak spectral phase
mismatch characterized by a zero-mean
residual phase distribution with variance
$\sigma_\phi^2$ over the occupied bandwidth,
after removing any physically irrelevant
global phase. Assuming that higher-order
phase cumulants are negligible, a
second-order cumulant expansion of the
coherent phase factor gives

\begin{equation}
 \mathcal{M}[\phi,\Delta] \approx \mathcal{M}[0,\Delta] e^{-\sigma_\phi^2/2}  \implies P_e \approx P_e^{(0)} e^{-\sigma_\phi^2}
\end{equation}
So a $10\%$ loss in $P_e$ corresponds to $\sigma_{\phi} \approx \sqrt{0.105}\approx0.32$ rad rms across the in-band frequencies. This sets a concrete specification for pixel-to-pixel SLM (spatial light modulator) errors and spectral-phase drift in a fielded link.

For a binary code with chips of width $\Omega=W/N_0$ and random sign errors of probability $p$, one obtains (Appendix \ref{app:effect})
\begin{equation}
\label{eq:phase}
\langle e^{i\phi} \rangle=(1-p)(+1)+p(-1)=1-2p  \implies P_e \approx P_e^{(0)} (1-2p)^2 
\end{equation}
so $p \leq 5\%$ keeps $P_e\geq0.81 P_e^{(0)}$.

\subparagraph{Time-resolved excitation under residual spectral-phase errors.}
To quantify the effect of imperfect decoding beyond its influence on the
maximum excitation, we examine the full time dependence of the atomic
response.  After applying the decoding mask, the photon has the residual
spectral phases $\phi_n=\tilde\theta_n-\theta_n$ on the $N_0=2N+1$
frequency chips.  For the rectangular spectrum considered above, its
normalized temporal envelope can be written as
\begin{equation}
\xi_{\rm in}(t;\{\phi_n\})
 = \sum_{n=-N}^{N}\xi_n(t)e^{i\phi_n},\qquad
\xi_n(t)=\frac{1}{N_0}\sqrt{\frac{W}{2\pi}}\,
  \mathrm{sinc}\left(\frac{\Omega t}{2}\right)
  e^{-in\Omega t}
\label{eq:decoded_temporal_profile}
\end{equation}
where frequencies
are measured relative to the atomic transition.  Perfect decoding
($\phi_n=0$ for every chip) recovers the unencoded temporal envelope.
Within the single-excitation, Markovian model of Appendix \ref{app:derivation}, the
excited-state probability for a fixed error realization is
\begin{equation}
P_e(t;\{\phi_n\})
=\left|\sum_{n=-N}^{N}e^{i\phi_n}c_n(t)\right|^2,
\qquad
c_n(t)=\sqrt{\gamma}\int_{-\infty}^{t}
 e^{-\gamma(t-\tau)/2}\xi_n(\tau)\,d\tau
\label{eq:phase_error_time_dynamics}
\end{equation}
where we take ideal coupling ($\beta=1$), as in the previous numerical
results.  This representation retains the temporal response of every
spectral chip, rather than replacing the effect of phase mismatch by a
single reduction factor.

To describe phase uncertainty, we take $\phi_n$ to be independent,
zero-mean Gaussian random variables with variance $\sigma_\phi^2$,
constant during an individual photon interaction.  The ensemble-averaged
excitation probability then follows exactly from
$\langle e^{i(\phi_n-\phi_m)}\rangle
=e^{-\sigma_\phi^2}$ for $n\ne m$
\begin{equation}
\langle P_e(t)\rangle_\phi
=e^{-\sigma_\phi^2}P_e^{(0)}(t)
+\left(1-e^{-\sigma_\phi^2}\right)
  \sum_{n=-N}^{N}|c_n(t)|^2
\label{eq:gaussian_phase_error_exact}
\end{equation}
where $P_e^{(0)}(t)=|\sum_n c_n(t)|^2$ is the perfectly decoded
excitation probability.  The second term is the finite-chip incoherent
contribution, absent from the coherent-sum approximation given above.
When the coherent contribution dominates, Eq. \eqref{eq:gaussian_phase_error_exact}
reduces to $\langle P_e(t)\rangle_\phi\simeq
P_e^{(0)}(t)e^{-\sigma_\phi^2}$.

\begin{figure}[t]
\centering
\includegraphics[width=0.7\linewidth]{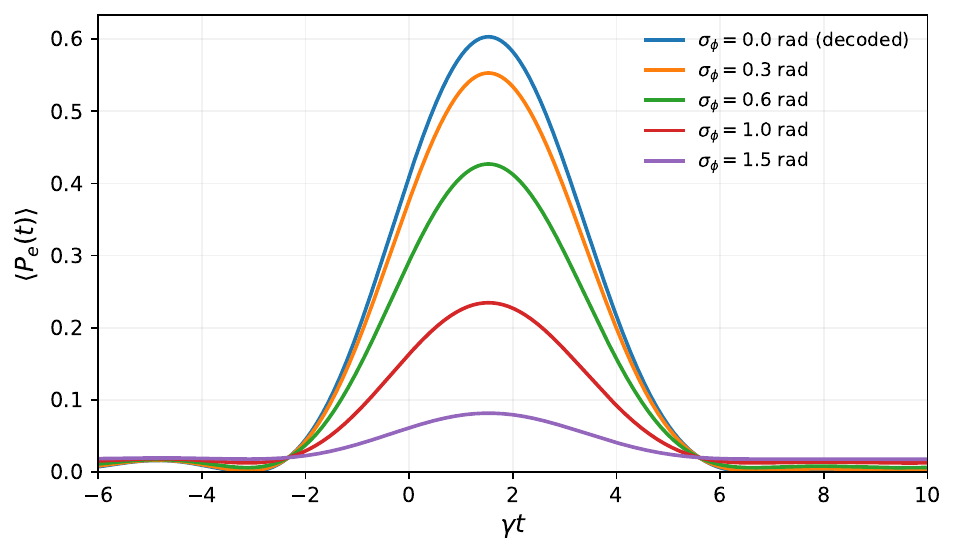}
\caption{Time-dependent ensemble-averaged atomic excitation probability
for an encoded single photon after imperfect spectral-phase decoding.
The independent residual phase on each of $N_0=31$ frequency chips is
Gaussian with zero mean and standard deviation $\sigma_\phi$.
Results use $W=1.5\gamma$ and $\beta=1$ and are calculated from
Eq. \eqref{eq:gaussian_phase_error_exact}.  The $\sigma_\phi=0$
curve corresponds to perfect decoding (the unencoded temporal mode).
The time origin is the center of the sinc envelope; the atomic
response includes the incident field at earlier times.
No additional environmental phase noise is assumed.}
\label{fig:residual_phase_dynamics}
\end{figure}

FIG. \ref{fig:residual_phase_dynamics} shows the resulting time-dependent
excitation for $N_0=31$ and $W=1.5\gamma$, varying the residual-phase
standard deviation from $0$ to $1.5$ rad.  The maximum excitation
falls from $0.603$ for perfect decoding to $0.553$, $0.427$, $0.235$,
and $0.082$ for $\sigma_\phi=0.3$, $0.6$, $1.0$, and $1.5$ rad,
respectively.  In particular, an rms error of $0.3$ rad reduces the
maximum excitation by approximately $8\%$, whereas a $1.0$ rad error
reduces it by approximately $61\%$.  Thus, receiver phase mismatch
not only suppresses the peak but also modifies the full excitation
trajectory through the relative contributions of the spectral chips.
These results demonstrate how residual spectral phase errors
modify the temporal evolution of the atomic excitation
probability and reduce the efficiency of photon absorption,
thereby quantifying the sensitivity of mode-selective
atom-photon interfaces to imperfect decoding in
quantum networks.

\subsection{Spectral CDMA and cross-talk in multiuser quantum networks}
\label{sec:cross_talk}
Recent developments in multiuser quantum networks
have highlighted the importance of spectral
and temporal control of single photons for
quantum communication
\cite{Huang2023,Myilswamy2025}.
In this context, spectral phase encoding
provides a means of distinguishing photons
addressed to different quantum-network nodes.
An important limitation in such networks is
cross-talk, which refers to the unintended
excitation of an atomic receiver by a photon
encoded for another user or node. This effect
arises when imperfect discrimination between
different spectral codes leaves a nonzero
overlap between a mismatched photon and the
atomic absorption mode. Consequently, the
degree of cross-talk is determined by the
spectral-code correlations and the
frequency-dependent response of the
receiving atom.

To quantify cross-talk arising from
spectral-code mismatch and evaluate the
selectivity of atomic receivers, we consider
$K$ users sharing a common optical frequency
band, each assigned a distinct binary
spectral phase code $\{\theta_{k,n}\}$
distributed over $N_0$ frequency chips.
To receive a photon from user $j$,
a node applies the compensating phase mask
$e^{i\theta_j(\omega)}$ before its
atom-photon interface. For the intended
photon ($k=j$), the encoder and decoder
phases cancel, allowing the spectral
components to contribute coherently to the
atomic excitation amplitude. In contrast,
for a photon from another user ($k\ne j$),
the residual chip phases
$\theta_{j,n}-\theta_{k,n}$ generally lead
to partial destructive interference,
suppressing the atomic excitation
probability and thereby reducing
interuser cross-talk.

To quantify this effect, define the complex
response weight of chip $n$ as
\begin{equation}
\alpha_n=\int_{\mathcal B_n}d\omega\,
\xi(\omega)\chi_A(\omega),
\qquad
\mathcal A_{k\to j}=
\sum_{n=-N}^{N}
\alpha_n e^{i(\theta_{j,n}-\theta_{k,n})}
\end{equation}
where $\chi_A(\omega)$ is the atomic response
defined in Eq. \eqref{eq:Lorentzian} and 
the amplitude $\mathcal A_{k\to j}$ is the
user-resolved extension of the excitation
amplitude introduced in Eq. \eqref{eq:amplitude}. It describes the
coupling of a photon transmitted by user
$k$ to the atomic receiver configured
for user $j$, accounting for the residual
spectral phase between their encoding and
decoding masks. For the correctly decoded
photon ($k=j$), these phases cancel,
yielding $\mathcal A_{j\to j}=\sum_n\alpha_n$.

If distinct users' binary codes are chosen
independently and uniformly, the residual
signs for $k\ne j$ are independent and
equally likely to be $+1$ or $-1$.
Consequently, the cross terms vanish
on averaging over codes, giving
\begin{equation}
\frac{
\mathbb{E}_{\rm codes}|\mathcal A_{k\to j}|^2
}{
|\mathcal A_{j\to j}|^2
}
=
\frac{\sum_n|\alpha_n|^2}{
|\sum_n\alpha_n|^2
}
\equiv\frac{1}{N_{\rm eff}}
\end{equation}
The quantity
$N_{\rm eff}=|\sum_n \alpha_n|^2/
\sum_n|\alpha_n|^2$
characterizes the number of effectively
phase-aligned spectral contributions at
the atomic receiver. When the chip weights
have approximately equal amplitudes and
phases, $N_{\rm eff}\simeq N_0$, so the
mean undesired coupling power is suppressed
by a factor $1/N_0$ relative to that of
the correctly decoded photon. Uneven
spectral weights, detuning, or phase
variation across the atomic response can
make $N_{\rm eff}$ smaller, emphasizing
that $1/N_0$ is an idealized scaling
rather than a universal bound for
every code realization.

For $K-1$ independently coded interfering
users with equal input strengths,
a corresponding average
signal-to-interference ratio (SIR) is
\begin{equation}
\mathrm{SIR}\simeq
\frac{N_{\rm eff}}{K-1}
\simeq\frac{N_0}{K-1}
\end{equation}
For example, $N_0=31$ and $K=5$ give
the illustrative flat-chip estimate
$\mathrm{SIR}\simeq31/4=7.75$.
Increasing the code length can therefore
improve user discrimination, while the
atomic linewidth and spectral mode matching
determine how much of that improvement is
available in practice. The statistical
derivation is presented in Appendix \ref{app:expectation}.
This estimate compares single-photon
coupling weights in the linear,
single-excitation description; simultaneous
multiphoton arrival, atomic saturation,
and nonideal hardware are not included.

\subsection{Mode-selective addressability in quantum networks}
Spectral phase encoding provides a mechanism
for controlling atom-photon interactions in
multiuser quantum networks through the
spectral-temporal structure of single-photon
wave packets. Unlike frequency-selective
addressing, which relies on assigning distinct
transition frequencies to different nodes,
the proposed approach allows users to share
the same optical bandwidth while being
distinguished by their spectral phase codes.
Each receiving node applies a user-specific
decoding mask, which determines the residual
spectral phase of the incoming photon and,
consequently, its overlap with the atomic
absorption mode.

For a correctly addressed photon, the
decoding operation compensates the imposed
spectral phases and restores the original
temporal wave packet. This enables the
frequency components to interfere coherently
in the atomic excitation amplitude, with
the achievable excitation probability
determined by the temporal-mode overlap
and the coupling efficiency $\beta$.
For a photon carrying a different spectral
code, the decoding operation leaves
nonzero residual phase differences.
These phases modify the interference
between spectral components contributing
to atomic excitation, generally reducing
the excitation probability and producing
the cross-talk suppression discussed above.
The atomic receiver therefore provides
mode-dependent excitation selectivity,
while the decoding mask determines which
encoded photon mode is preferentially
coupled to the atomic transition.

An important consequence is that individual
atomic nodes can be addressed through
spectral-mode discrimination without
requiring distinct atomic resonance
frequencies. The effectiveness of this
mechanism depends on the spectral-code
correlations, the atomic linewidth, the
accuracy of phase compensation, and the
coupling efficiency. Although sufficiently
distinct codes can suppress unwanted
excitation, finite spectral overlap and
experimental imperfections prevent perfect
isolation between users. Moreover,
mode-selective excitation alone does not
guarantee secure communication, since a
photon prepared with the appropriate
spectral code can reproduce the intended
coupling response. Nevertheless, spectral
phase encoding offers a physically
motivated approach to reducing cross-talk
and improving node addressability in
shared-bandwidth quantum networks.

\subsection{Design considerations and experimental implementation}
The preceding analysis establishes several
physical requirements for the implementation
of spectrally encoded atom-photon interfaces.
A central consideration is the matching
between the photon bandwidth $W$ and the
atomic linewidth $\gamma$, which determines
the balance between temporal localization
and resonant spectral overlap. For the
rectangular single-photon spectrum considered
here, the maximum excitation occurs at
$W/\gamma\approx 1.39$, as derived in
Appendix \ref{app:optimal}. This condition
is specific to the adopted photon waveform
and atomic response. More generally,
efficient single-photon absorption depends
on matching the incident temporal mode
to the atomic excitation dynamics \cite{Brecht,Aljunid,Wang2}.
The achievable excitation is further
limited by the optical coupling efficiency
$\beta$, which can be improved through
appropriate spatial mode matching and
engineered light-matter interfaces \cite{Lindlein,Rephaeli,Chen}.

For multiuser operation, the spectral
code length $N_0$ determines the number
of independently controlled frequency
components and influences discrimination
between different encoded photon modes.
Spectral phase encoding has been developed
in optical code-division multiple-access
systems, while recent quantum-photonic
studies have demonstrated the importance
of spectral and temporal mode control
for quantum communication\cite{Huang2023,Myilswamy2025}. As discussed in Section \ref{sec:cross_talk},
the mean cross-talk between independently
assigned binary codes decreases
approximately as $N_0^{-1}$ for
comparably weighted spectral chips.
However, increasing $N_0$ also places
greater demands on spectral resolution
and phase calibration. Within the
weak-phase-error approximation of
Section \ref{sec:tolerance}, the relative
excitation probability follows
$P_e/P_e^{(0)}\simeq
\exp(-\sigma_\phi^2)$.
An rms residual phase error
$\sigma_\phi\lesssim 0.3$ rad therefore
corresponds to an excitation reduction
below approximately $10\%$ under these
assumptions. Frequency detuning must
also be controlled by matching the
photon carrier frequency to the atomic
transition, while temporal misalignment
can be compensated through a linear
spectral phase correction.

An experimental realization could combine
a heralded single-photon source with
programmable spectral phase modulation.
The photon would first pass through
a spectral pulse shaper implementing
the encoding mask $\theta(\omega)$,
followed by propagation through an
optical communication channel.
At the receiving node, a second
spectral modulator would implement
the decoding operation
$D(\omega)=e^{i\tilde{\theta}(\omega)}$,
restoring the desired photon mode
when the encoding and decoding masks
are matched. Such frequency-domain
control is consistent with established
spectral-encoding techniques and
developments in quantum frequency-bin
photonics
\cite{Huang2023,Myilswamy2025}.
The decoded photon would subsequently
interact with a two-level emitter
through a suitable optical interface.
Single-atom excitation in free space
and coherent photon absorption in
one-dimensional photonic structures
provide relevant physical foundations
for this arrangement
\cite{Aljunid,Rephaeli,Chen}.

The theoretical predictions could be
examined by systematically varying
the bandwidth ratio $W/\gamma$,
the spectral code length $N_0$,
and the residual phase-error variance
$\sigma_\phi^2$, while independently
characterizing the coupling efficiency
$\beta$. Measurements of atomic
excitation for matched, imperfectly
decoded, and mismatched photons would
allow the influence of spectral phase
compensation to be distinguished from
the limitations imposed by the atomic
response. In particular, the dependence
of excitation on phase mismatch and
the predicted $N_0^{-1}$ suppression
of ensemble-averaged cross-talk could
be tested experimentally. These
measurements would establish the
operating conditions under which
spectral encoding provides effective
mode discrimination and selective
atom-photon coupling in multiuser
quantum networks.

\section{Conclusion}
\label{sec:conclusion}
We investigated spectral phase encoding as a mechanism
for controlling the excitation of two-level atomic
receivers by single-photon wave packets in quantum
networks. Within the Heisenberg-Langevin description,
the atomic response is governed by the temporal-mode
overlap between the incident photon and the
time-reversed spontaneous-emission mode. Spectral
encoding preserves the photon number and spectral
intensity while modifying the interference between
frequency components. The resulting excitation therefore
depends on the phase sequence and its overlap with the
atomic response, rather than on code length alone.
Matched decoding restores the original photon mode
and its excitation probability, although it does not
necessarily produce the optimal absorption waveform.

At the network level, this dependence provides a basis
for distinguishing photons addressed to different nodes
within a shared optical bandwidth. We derived the
average cross-talk for independently assigned binary
codes and identified how unequal spectral weights and
phase variations across the atomic response limit
user discrimination. The favorable suppression with
increasing code length requires sufficiently comparable
and phase-aligned spectral contributions. We also
derived the time-dependent response under independent
Gaussian decoding errors, retaining the incoherent
contribution that remains when coherent interference
is suppressed. These results connect spectral-code
design and decoding accuracy to the excitation
selectivity of an atomic receiver.

Several limitations remain important for interpreting
these results. The analysis assumes a two-level emitter
with Markovian decay and treats excitation within the
single-excitation manifold. It therefore does not
describe simultaneous multiphoton arrivals, atomic
saturation, or the full operation of a multiuser
network. Moreover, selective transient excitation
does not by itself establish reliable quantum-state
storage or transfer: these tasks require control of
the subsequent atomic dynamics and preservation of
the encoded quantum information. Suppression averaged
over random codes also does not guarantee low
cross-talk for every assigned code pair.

An experimental test could combine a heralded
single-photon source, programmable spectral phase
encoding and decoding, and an atomic interface with
independently characterized coupling efficiency.
The principal challenges include resolving spectral
chips within the atomic linewidth, limiting optical
loss, and maintaining phase calibration and temporal
stability. Measurements of the time-dependent atomic
excitation under matched and mismatched decoding
would test mode selectivity, while controlled phase
perturbations would probe decoding tolerance.
Characterizing photon transmission and the incident
temporal mode would be necessary to distinguish
imperfect phase compensation from optical loss
and spatial coupling limitations.

Future works should optimize spectral codes jointly
with the atomic response and the receiver's temporal
detection window, including bounds on the largest
cross-talk between assigned users. Other directions
include correlated phase errors, propagation-induced
dispersion, finite spectral resolution, and the
trade-off between code length and communication rate.
Extending the treatment to multiple incident photons,
multilevel emitters, and controlled transfer into
long-lived states would connect excitation selectivity
to practical quantum-memory operation. Such studies
would establish how spectral phase control can
contribute to addressable atom-photon interfaces
under realistic network conditions.

\subparagraph{Acknowledgments}
The author gratefully acknowledges Jawad Salehi,
Ali Soltanmanesh and Mohammad Rezai for their
valuable discussions and constructive comments.

\subparagraph{Data availability}
The numerical data supporting the findings
of this study are available from the
corresponding author upon reasonable request.

\appendix

\section{Derivation of the atomic excitation overlap bound}
\label{app:derivation}
We derive the temporal-mode overlap bound in
Eq. \eqref{eq:bound} using the same atom-photon
interaction model introduced in
Section \ref{sec:model}. The Heisenberg-Langevin
equations describe the atomic dynamics through
the evolution of its operators. However, when
the atom is initially in its ground state and
the incident field contains a single photon,
the rotating-wave approximation restricts
the dynamics to the single-excitation manifold.
In this subspace, the atomic excitation can
equivalently be described by the excited-state
probability amplitude $c_e(t)$. In particular,
the atomic inversion satisfies
$\langle\sigma_z(t)\rangle=2|c_e(t)|^2-1$,
so that
\begin{equation}
P_e(t)=\frac{1+\langle\sigma_z(t)\rangle}{2}
      =|c_e(t)|^2
\end{equation}
This is precisely the excitation probability
defined in Eq. \eqref{eq:probability}, rather
than a new definition. The single-excitation
representation therefore provides a direct
way to derive the optimal absorption mode
from the same underlying atomic dynamics.

We denote the total atomic decay rate by
$\gamma$, consistently with Section \ref{sec:model},
and the coupling rate into the incident
optical channel by
$\gamma_{\mathrm{in}}=\beta\gamma$,
where $0\leq\beta\leq 1$.
The remaining decay channels contribute
$\gamma-\gamma_{\mathrm{in}}$ to the total
linewidth. Within the Markov approximation,
the excited-state amplitude satisfies
\begin{equation}
\dot c_e(t)
=-\frac{\gamma}{2}c_e(t)
+\sqrt{\beta\gamma}\,\xi(t)
\label{eq:ce_dyn}
\end{equation}
where $\xi(t)$ is the temporal amplitude of
the incident single photon. Equation
\eqref{eq:ce_dyn} is the single-excitation
form of the dynamics underlying the
Heisenberg-Langevin equations
\eqref{eq:sigmaminus} and \eqref{eq:sigmaz}.

Assuming that the atom is initially in
its ground state in the remote past,
$c_e(-\infty)=0$, the formal solution is
\begin{equation}
c_e(t)=\sqrt{\beta\gamma}
\int_{-\infty}^{t}
e^{-\gamma(t-\tau)/2}
\xi(\tau)\,d\tau
\label{eq:ce_conv}
\end{equation}
The incident single-photon temporal mode
is normalized according to
\begin{equation}
\int_{-\infty}^{\infty}
|\xi(\tau)|^2\,d\tau=1
\label{eq:norm}
\end{equation}
Eq. \eqref{eq:ce_conv} shows that
atomic excitation results from the coherent
accumulation of the incident photon amplitude,
weighted by the exponentially decaying
atomic response. The contributions arriving
at different times interfere in the
excited-state amplitude, while earlier
contributions are suppressed by spontaneous
emission.

To identify the photon mode producing
the largest excitation at a given time
$t$, we introduce the normalized atomic
absorption mode
\begin{equation}
\xi_{\mathrm{opt},t}(\tau)
=\sqrt{\gamma}\,
e^{\gamma(\tau-t)/2}\Theta(t-\tau)
=\xi_{\mathrm{opt}}(\tau-t)
\label{eq:optimal_shifted}
\end{equation}
where
\begin{equation}
\xi_{\mathrm{opt}}(t)
=\sqrt{\gamma}\,
e^{\gamma t/2}\Theta(-t)
\label{eq:optimal_mode_app}
\end{equation}
This is the time-reversed spontaneous-emission
mode introduced in
Eq. \eqref{eq:temporal_profile}.
Its normalization follows directly from
\begin{equation}
\int_{-\infty}^{\infty}
|\xi_{\mathrm{opt},t}(\tau)|^2\,d\tau
=\gamma\int_{-\infty}^{t}
e^{\gamma(\tau-t)}\,d\tau=1
\end{equation}
Using Eq. \eqref{eq:optimal_shifted},
the excited-state amplitude in
Eq. \eqref{eq:ce_conv} can be written
as the temporal-mode overlap
\begin{equation}
\begin{aligned}
c_e(t)
&=\sqrt{\beta}
\int_{-\infty}^{\infty}
\xi_{\mathrm{opt},t}^{*}(\tau)
\xi(\tau)\,d\tau\\
&=\sqrt{\beta}\,
\langle\xi_{\mathrm{opt},t}|\xi\rangle
\end{aligned}
\label{eq:ce_overlap}
\end{equation}
Consequently, the excitation probability
becomes
\begin{equation}
\begin{aligned}
P_e(t)
&=\beta\left|
\langle\xi_{\mathrm{opt},t}|\xi\rangle
\right|^2\\
&\leq\beta\,
\langle\xi_{\mathrm{opt},t}|
\xi_{\mathrm{opt},t}\rangle
\langle\xi|\xi\rangle\\
&=\beta
\end{aligned}
\label{eq:CS_bound}
\end{equation}
where the inequality follows from the
Cauchy-Schwarz inequality and the
normalization of both temporal modes.

Maximizing over the observation time
therefore gives
\begin{equation}
P_e^{\max}
=\beta\max_t
\left|
\int_{-\infty}^{\infty}
d\tau\,\xi^*(\tau)
\xi_{\mathrm{opt}}(\tau-t)
\right|^2
\leq\beta
\label{eq:overlap_bound}
\end{equation}
This reproduces Eq.~\eqref{eq:bound}
of the main text. Equality is attained
when the incident single-photon temporal
mode matches the time-shifted optimal
absorption mode up to a global phase.
Physically, the exponentially rising
incident waveform compensates for the
natural atomic decay and maximizes the
constructive accumulation of excitation
amplitude. The coupling efficiency
$\beta$ determines the maximum achievable
population even under perfect temporal
mode matching.

For completeness, if the atom-photon
interaction is instead initialized at a
finite time $t_0$, with $c_e(t_0)=0$,
the same dynamical equation gives
\begin{equation}
c_e(t)=\sqrt{\beta\gamma}
\int_{t_0}^{t}
e^{-\gamma(t-\tau)/2}
\xi(\tau)\,d\tau,
\qquad t\geq t_0
\label{eq:ce_finite_time}
\end{equation}
In this case, only the portion of the
incident wave packet interacting with
the atom after $t_0$ contributes to
the excitation. The bound
$P_e(t)\leq\beta$ remains valid, but
the overlap must be evaluated over the
corresponding interaction interval.
Thus, the finite-time and remote-past
initial conditions describe different
temporal preparation protocols within
the same atom-photon interaction model.

\section{Influence of binary phase-flip errors on atomic excitation}
\label{app:effect}
We divide the occupied bandwidth $W$ into
$N_0$ spectral chips of width $\Omega=W/N_0$.
Each chip contributes a complex excitation
amplitude determined by the photon spectrum
and the atomic response. Under ideal decoding,
the imposed spectral phases are compensated,
allowing these amplitudes to interfere
coherently. Denoting the contribution of
the $n$th chip by $a_n$, the total excitation
amplitude is $A^{(0)}=\sum_{n=1}^{N_0}a_n$,
with $P_e^{(0)}\propto |A^{(0)}|^2$.
Binary phase-flip errors reverse the signs
of selected contributions, modifying their
interference and potentially suppressing
atomic excitation.

We suppose a binary phase error model: on each chip $n$, a random sign flip occurs with probability $p$, i.e.
\bq
\phi_n=\begin{cases}
0 & \text{w.p. } 1-p,\\
\pi & \text{w.p. } p,
\end{cases}
\qquad X_n \equiv e^{i\phi_n}\in\{+1,-1\}
\eq
independently across $n$. Then
\begin{equation}
\langle e^{i\phi_n}\rangle=\langle X_n\rangle=(1-p)(+1)+p(-1)=1-2p
\label{eq:mean_X}
\end{equation}
With errors, the excitation amplitude is
\begin{equation}
A=\sum_{n=1}^{N_0} a_n X_n,
\qquad
\Rightarrow\qquad
\langle A\rangle=(1-2p)\sum_{n=1}^{N_0} a_n=(1-2p)\,A^{(0)}
\label{eq:mean_A}
\end{equation}

\paragraph*{Exact interference-power expression.}
Because $X_n$ are independent, $X_n^2\equiv 1$, and $\langle X_n X_m\rangle=\langle X_n\rangle\langle X_m\rangle=(1-2p)^2$ for $n\neq m$,
\begin{align}
\left\langle |A|^2 \right\rangle
&= \sum_{n=1}^{N_0} |a_n|^2 \langle X_n^2\rangle
+ \sum_{\substack{n,m=1\\ n\ne m}}^{N_0} a_n a_m^* \langle X_n X_m\rangle \nonumber\\
&= \sum_{n=1}^{N_0} |a_n|^2 + (1-2p)^2 \sum_{\substack{n,m=1\\ n\ne m}}^{N_0} a_n a_m^* \nonumber\\
&= (1-2p)^2 \Big|\sum_{n=1}^{N_0} a_n\Big|^2
\;+\; \big[1-(1-2p)^2\big]\sum_{n=1}^{N_0} |a_n|^2 \nonumber\\
&= (1-2p)^2 |A^{(0)}|^2 \;+\; 4p(1-p)\sum_{n=1}^{N_0} |a_n|^2
\label{eq:exact_power}
\end{align}
Thus, the \emph{exact} mean excitation probability is
\begin{equation}
\langle P_e\rangle \ \propto\ (1-2p)^2\,|A^{(0)}|^2 \;+\; 4p(1-p)\sum_{n=1}^{N_0} |a_n|^2
\end{equation}

\paragraph*{Coherent-sum (large-$N_0$) approximation.}
When the decoded chips add coherently (e.g., $a_n\approx a$) we have
$|A^{(0)}|^2=N_0^2|a|^2$ and $\sum_n |a_n|^2=N_0|a|^2$, so the second term in
\eqref{eq:exact_power} is smaller by a factor $\sim 1/N_0$:
\begin{equation}
\frac{4p(1-p)\sum_n |a_n|^2}{(1-2p)^2 |A^{(0)}|^2}
= \frac{4p(1-p)}{(1-2p)^2}\cdot \frac{1}{N_0}\ \ll\ 1
\qquad (N_0\ \text{large})
\end{equation}
Hence
\begin{equation}
P_e \ \approx\ P_e^{(0)}\,(1-2p)^2
\label{eq:approx_Pe}
\end{equation}
which is the stated result in Eq. \eqref{eq:phase}. Eq. \eqref{eq:approx_Pe} follows directly from the mean chip factor \eqref{eq:mean_X} and the dominance of the coherent term for large $N_0$; the exact correction is given by \eqref{eq:exact_power}.

\section{Statistical analysis of spectral-code cross-talk and signal-to-interference ratio}
\label{app:expectation}
We split the occupied band into $N_0$ chips $\{\omega_n\}_{n=1}^{N_0}$. 
Each user $k$ employs a phase code $\theta_k(\omega_n)$. 
We average over i.i.d.\ random codes drawn from the binary ensemble
\begin{equation}
\mathcal{E}_{\mathrm{bin}}:\quad 
\theta_k(\omega_n)\in\{0,\pi\}\ \text{i.i.d. with}\ 
\Pr[\theta_k(\omega_n)=0]=\Pr[\theta_k(\omega_n)=\pi]=\tfrac12
\end{equation}
and denote expectations by $\mathbb{E}_{\mathrm{codes}}[\cdot]$.
For two users $j\neq k$, we define the chipwise code-difference factor
\begin{equation}
X_n \equiv e^{i[\theta_j(\omega_n)-\theta_k(\omega_n)]}\in\{+1,-1\}
\end{equation}
Under $\mathcal{E}_{\mathrm{bin}}$, $\{X_n\}$ are i.i.d. Rademacher variables with
\begin{equation}
\mathbb{E}[X_n]=0,\qquad \mathbb{E}[X_n^2]=1
\end{equation}
The normalized discrete code correlation that appears in overlap integrals is
\begin{equation}
C_{jk}\ \equiv\ \frac{1}{N_0}\sum_{n=1}^{N_0} X_n
\ =\ \frac{1}{N_0}\sum_{n=1}^{N_0} e^{i[\theta_j(\omega_n)-\theta_k(\omega_n)]}
\end{equation}
By linearity and independence,
\begin{align}
\mathbb{E}_{\mathrm{codes}}[C_{jk}] 
&= \frac{1}{N_0}\sum_{n=1}^{N_0}\mathbb{E}[X_n] = 0\\
\mathrm{Var}_{\mathrm{codes}}(C_{jk})
&= \mathbb{E} \left[ \left| \frac{1}{N_0}\sum_{n=1}^{N_0} X_n \right|^2 \right]
= \frac{1}{N_0^2}\sum_{n=1}^{N_0}\mathbb{E}[X_n^2]
= \frac{1}{N_0}
\end{align}
Hence $C_{jk}\xrightarrow{\mathbb{P}}0$ as $N_0\to\infty$, and typical cross-talk scales as $N_0^{-1/2}$ in amplitude (or $N_0^{-1}$ in power).

Let $a_n$ denote the deterministic complex amplitude on chip $n$ for the desired user $j$ after decoding (it can collect spectral weight, atomic response, and any fixed factors). 
The interference at the node from another user $k\neq j$ is
\begin{equation}
I_{jk}\ \equiv\ \sum_{n=1}^{N_0} a_n\,X_n
\end{equation}
Then
\begin{equation}
\mathbb{E}_{\mathrm{codes}}[I_{jk}] = 0,
\qquad
\mathbb{E}_{\mathrm{codes}} \big[\,|I_{jk}|^2\,\big]
= \sum_{n=1}^{N_0} |a_n|^2
\end{equation}
since cross terms vanish by independence and $\mathbb{E}[X_n X_m]=0$ for $n\neq m$.
With $K-1$ independent interfering users,
\begin{equation}
\mathbb{E}_{\mathrm{codes}} \left[ \left| \sum_{k\neq j} I_{jk} \right|^2 \right]
= (K-1)\sum_{n=1}^{N_0}|a_n|^2
\end{equation}
The desired (decoded) field amplitude for user $j$ is
\begin{equation}
S_j\ \equiv\ \sum_{n=1}^{N_0} a_n
\end{equation}
so a natural signal-to-interference ratio (SIR) is
\begin{equation}
\mathrm{SIR}\ \equiv\ 
\frac{|S_j|^2}{\mathbb{E}_{\mathrm{codes}} \left[\sum_{k\neq j}|I_{jk}|^2\right]}
= \frac{\left|\sum_{n=1}^{N_0} a_n\right|^2}{(K-1)\sum_{n=1}^{N_0}|a_n|^2}
\label{eq:SIR_general}
\end{equation}
For flat chips $a_n=a$,
\begin{equation}
|S_j|^2 = |a|^2 N_0^2,\qquad \sum_{n}|a_n|^2 = |a|^2 N_0
\quad\Rightarrow\quad
\ \mathrm{SIR}=\dfrac{N_0}{K-1}\
\end{equation}

\section{Optimal bandwidth for single-photon atomic excitation}
\label{app:optimal}
The excitation probability of a two-level atom driven by a single photon with normalized spectral amplitude $\xi(\omega)$
is proportional to the squared overlap between $\xi(\omega)$ and the atomic susceptibility $\chi_A(\omega)$
\begin{equation}
P_e(W)\ \propto\ \Big| \int_{-\infty}^{\infty} \xi(\omega)\, \chi_A(\omega)\, d\omega \Big|^2 
\label{eq:Pe_def}
\end{equation}
Considering a rectangular photon spectrum of total width $W$ centered at the atomic transition frequency $\omega_A$, and the frequency-domain response $\chi_A(\omega)$ in Eq. \eqref{eq:Lorentzian}, Eq. \eqref{eq:Pe_def} gives
\begin{equation}
\mathcal{M}(W)
= \frac{\sqrt{\gamma}}{\sqrt{W}}
\int_{-W/2}^{W/2}
\frac{d\omega}{\gamma/2 - i\omega}
= \frac{2\sqrt{\gamma}}{\sqrt{W}}
\tan^{-1} \left(\frac{W}{\gamma}\right)
\label{eq:overlap_W}
\end{equation}
where $\mathcal{M}(W)$ represents the (dimensionless) spectral overlap amplitude.
The excitation probability then scales as
\begin{equation}
P_e(W)\ \propto\ \frac{\gamma}{W}
\left[\tan^{-1} \left(\frac{W}{\gamma}\right)\right]^2
\label{eq:Pe_vs_W}
\end{equation}

The behavior of Eq. \eqref{eq:Pe_vs_W} can be analyzed in the two limiting regimes
\begin{align}
P_e(W) &\propto W, \qquad W\ll\gamma
\label{eq:D4}\\
P_e(W) &\propto \frac{1}{W}, \qquad W\gg\gamma
\label{eq:D5}
\end{align}
Therefore, $P_e(W)$ must reach a maximum for an intermediate bandwidth.
Setting $\partial P_e/\partial W=0$ in Eq. \eqref{eq:Pe_vs_W} yields a transcendental equation
\begin{equation}
\tan^{-1} \left(W/\gamma\right)=2(W/\gamma)/(1+(W/\gamma)^2)
\end{equation}
whose numerical solution is $W/\gamma\simeq1.39$. Physically, this condition corresponds to matching the photon’s spectral width to the atomic linewidth:
if $W\ll\gamma$, the pulse is too long in time and couples weakly,
whereas for $W\gg\gamma$ much of the photon energy lies off-resonance and is not absorbed.
The optimal intermediate regime ensures efficient temporal and spectral overlap between the encoded photon and the atomic transition.

\bibliographystyle{unsrt}
\bibliography{Ref}
\end{document}